\documentclass{cernrep}

\usepackage{texnames}
\usepackage[T1]{fontenc}

\usepackage{endnotes}       
\usepackage{color}
\def\rmi{\mathrm{i}} 
\def\rmd{\mathrm{d}} 
\def\rme{\mathrm{e}} 
\def\u{\:}           
\usepackage{amsfonts}
 \DeclareSymbolFont{upgreek}{U}{eur}{m}{n}
 \DeclareMathSymbol{\rmalpha}{0}{upgreek}{"0B}
 \DeclareMathSymbol{\rmbeta}{0}{upgreek}{"0C}
 \DeclareMathSymbol{\rmgamma}{0}{upgreek}{"0D}
 \DeclareMathSymbol{\rmdelta}{0}{upgreek}{"0E}
 \DeclareMathSymbol{\rmmu}{0}{upgreek}{"16}
 \DeclareMathSymbol{\rmpi}{0}{upgreek}{"19}
 \DeclareMathSymbol{\rmsigma}{0}{upgreek}{"1B}
 \DeclareMathSymbol{\rmphi}{0}{upgreek}{"1E}
 \DeclareMathSymbol{\rmomega}{0}{upgreek}{"21}

\begin{document}

\title{Electron Beam Ion Sources}

\author{G.Zschornack$^{a,b}$, M.Schmidt$^b$ and A.Thorn$^b$}

\institute{$^a$University of Technology Dresden, Helmholtz-Zentrum Dresden-Rossendorf, Dresden, Germany \newline
$^b$Dreebit GmbH, Dresden, Germany}

\maketitle

\begin{abstract}
Electron beam ion sources (EBISs) are ion sources that work based on the principle of electron impact ionization, allowing the production of very highly charged ions. The ions produced can be extracted as a DC ion beam as well as ion pulses of different time structures. In comparison to most of the other known ion sources, EBISs feature ion beams with very good beam emittances and a low energy spread. Furthermore, EBISs are excellent sources of photons (X-rays, ultraviolet, extreme ultraviolet, visible light) from highly charged ions. This chapter gives an overview of EBIS physics, the principle of operation, and the known technical solutions. Using examples, the performance of EBISs as well as their applications in various fields of basic research, technology and medicine are discussed.
\end{abstract}

\section{Introduction}

The idea to develop electron beam ion sources (EBISs) was constituted by the need for multiply charged ions for accelerator applications to derive high final particle kinetic energies and by a general scientific interest to study exotic states of matter as represented by highly charged ions (HCIs).

\begin{figure}[htbp!]
\centering\includegraphics[width=0.475\linewidth]{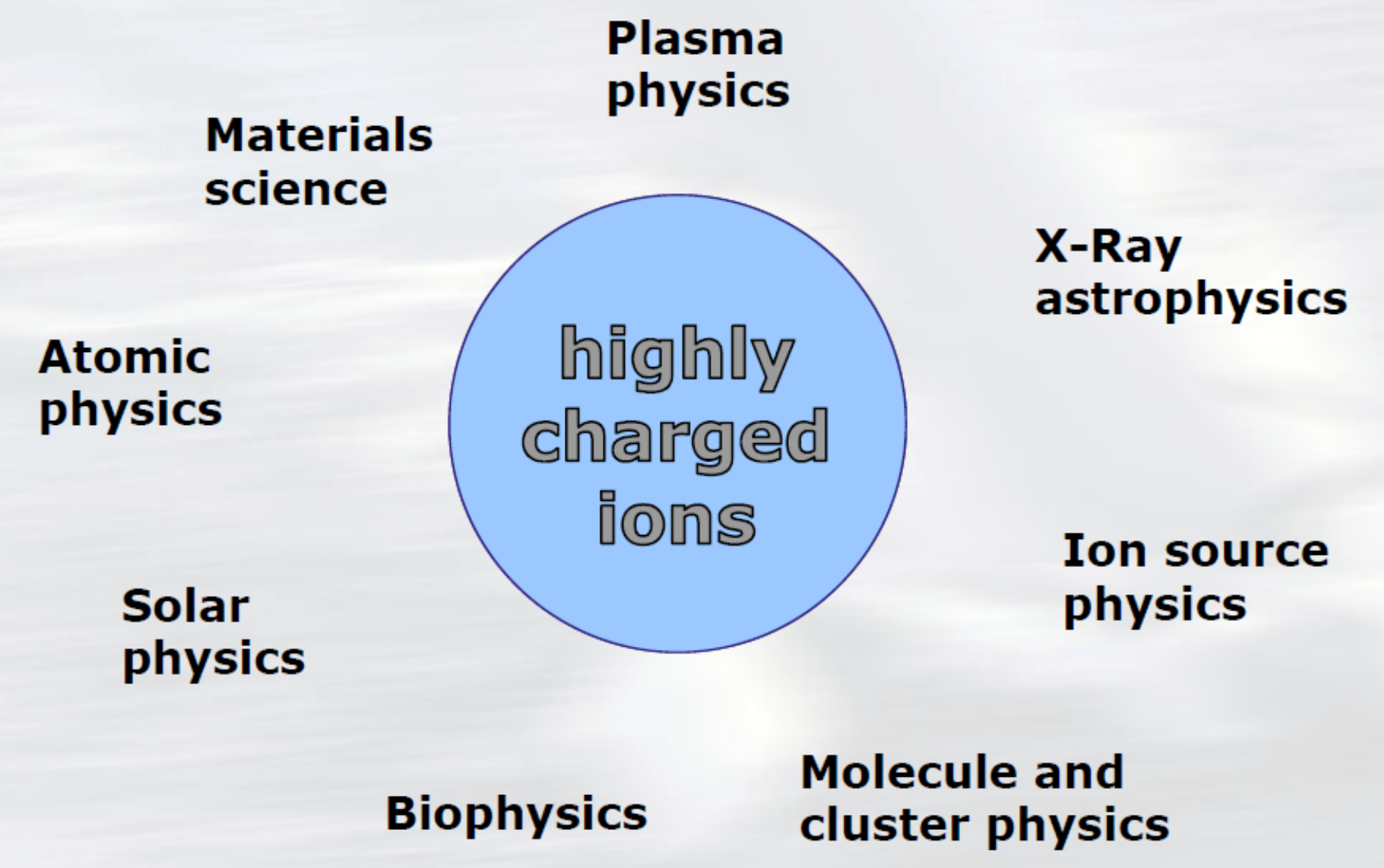}
\hspace{3mm}
\includegraphics[width=0.480\linewidth]{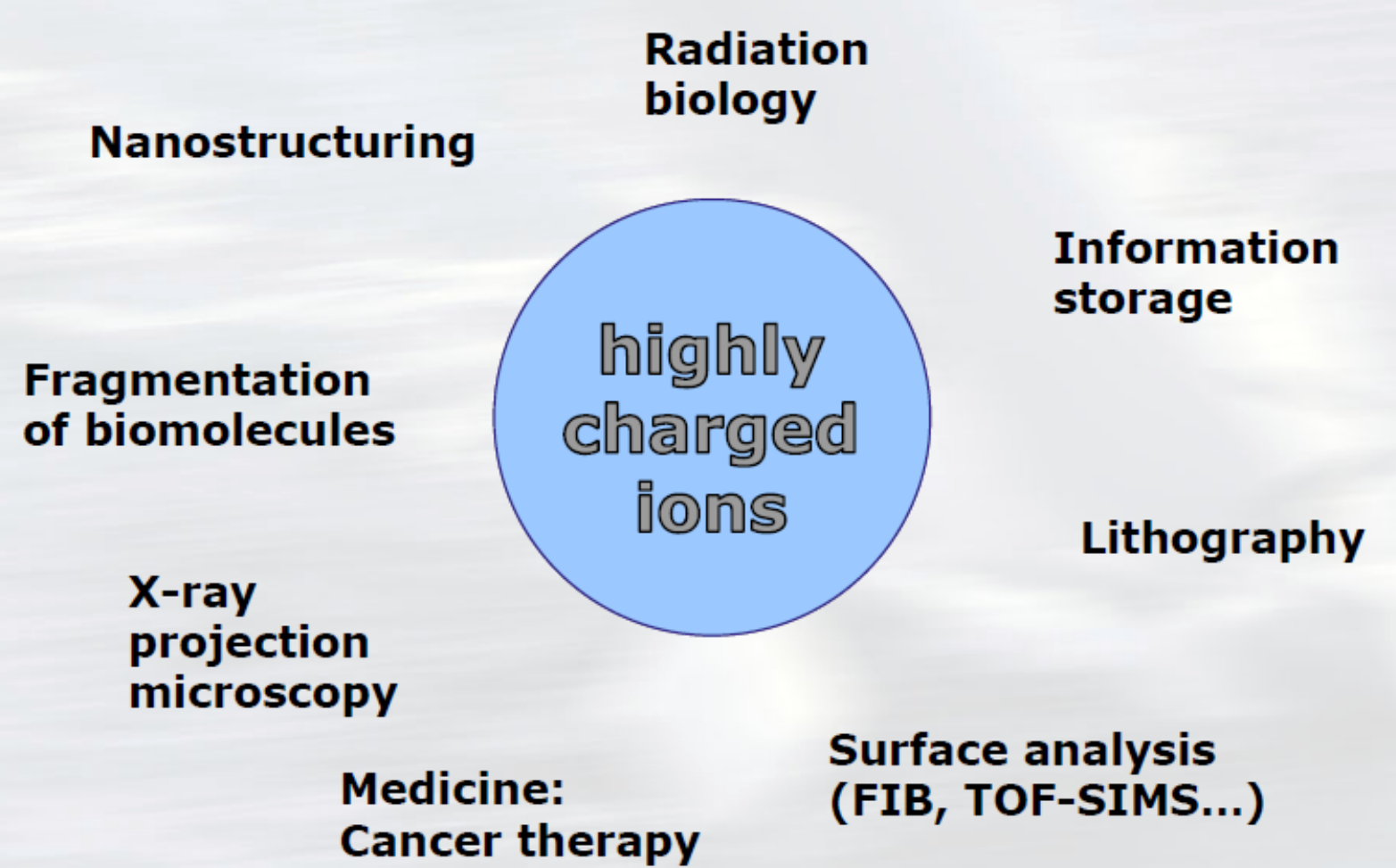}
\caption{Some applications of HCIs in basic research (left side) and possible applications in technology (right side)}
\label{fig:applications}
\end{figure}

Highly charged ions possess properties that differ in many ways from atoms and low charged ions. The following are some of the distinguishing characteristics.
\begin{itemize}
\item Due to the ionization process, a large amount of potential energy is stored in HCIs. For example, a Xe$^{44+}$ ion has a potential energy of about 51{\u}keV.
\item During ion--surface interactions, this potential energy is released within a few 10{\u}fs over an area of about 100{\u}nm$^2$, which leads to power densities of 10$^{12}$ up to 10$^{14}${\u}W{\u}cm$^{-2}$.
\item Due to the high power density in the surface interaction area compared to low charged ions (LCIs), HCIs produce high amounts of secondary particles such as neutrals as well as secondary electrons and secondary ions.
\item HCIs possess very strong static electric fields, reaching from 10$^{14}$ up to 10$^{16}${\u}V{\u}cm$^{-1}$.
\item HCIs are characterized by a very effective stopping power (e.g.\ for Au$^{69+}$, 100{\u}keV{\u}nm$^{-1}$).
\item Due to their high ionic charge, HCIs can be accelerated very effectively ($\propto q$ for linear accelerators and $\propto q^2$ for circular accelerators; $q$ denotes the ion charge state).
\item Nowadays, HCIs can be produced in a compact laboratory set-up. Large accelerator structures are used if high HCI currents are necessary for a specific experiment but are not required for many applications.
\end{itemize}
These unique properties are the reason why the interest in HCIs is constantly growing and the number of applications increases. Figure \ref{fig:applications} gives several examples for applications of HCIs in basic research as well as technology.

HCIs appear only in the plasma of fusion facilities, the corona of the Sun, at the border of black holes, and in similar exotic cosmic regions. The first publications on HCIs date back more than 80~years and the number of publications has increased continually. In the first classical papers Bowen and Millikan reported on the production of Se$^{20+}$ ions in an arc discharge \cite{bib:Bowen1925} in 1925 and Edlen explained the origin of spectral lines in the corona of the Sun as emission lines from 10- to 15-fold ionized iron, calcium and nickel atoms in 1942 \cite{bib:Edlen1942}.

A classical approach to produce HCIs in the laboratory is the use of ion accelerators by stripping of energetic LCIs to all practically possible ion charge states. As a result of the stripping process, HCIs of high kinetic energy are produced, which must be decelerated in many cases for further use. Another way to access HCIs is the use of powerful ion sources such as the following:
\begin{itemize}
\item EBISs;
\item electron cyclotron resonance (ECR) ion sources (ECRISs) (for basics see Ref.~\cite{bib:Geller1996}); or
\item laser ion sources (for basics see Ref.~\cite{bib:Brown2004}).
\end{itemize}

This present chapter is about EBISs. For details on other kinds of ion sources, we refer to the citations above and the references therein. The aim of this work is to give a short up-to-date overview about the function and use of EBIS technology. More details can be found in the literature, e.g.\ in the book of Gillaspy \cite{bib:Gillaspy2001} and the paper written by Currell \cite{bib:Currell2003}. More detailed information about the basic physics of HCIs can be found in the review by Gillaspy \cite{bib:Gillaspy2001a} and the book written by Beyer and Shevelko \cite{bib:Beyer2003}.

\section{EBIS: the basic idea}

\subsection{The technical solution}

\begin{figure}[htbp!]
\centering\includegraphics[width=0.7\linewidth]{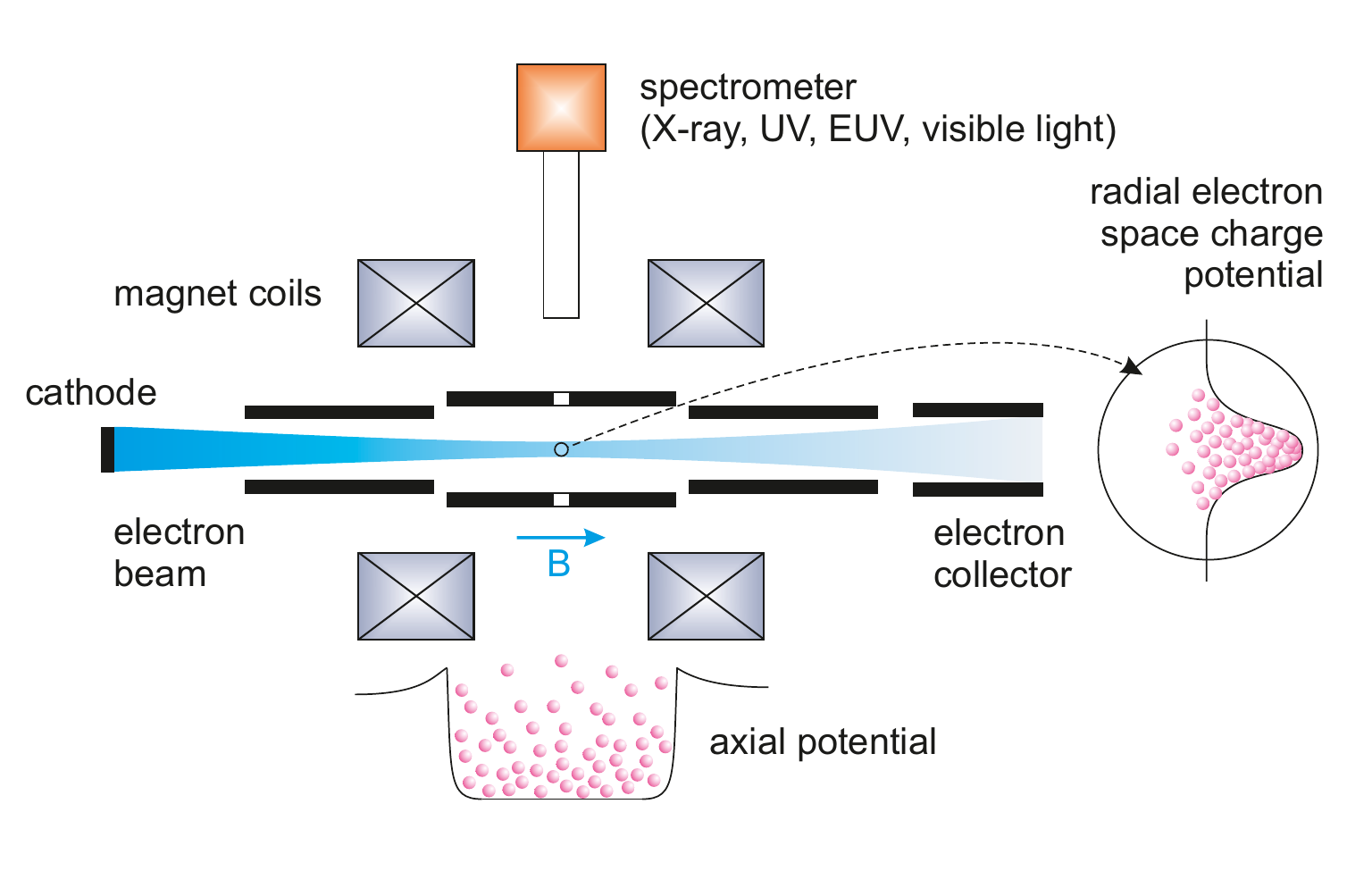}
\caption{Principle of operation of an EBIS}
\label{fig:EBIS-Funktionsprinzip}
\end{figure}

EBIS/T ion sources have been known for more than 40~years and the technical design of these sources has been continually improved with increasing technical possibilities. The principle of operation of an EBIS is shown in Fig.~\ref{fig:EBIS-Funktionsprinzip}.

The ion production in an EBIS is based on electron impact ionization in a high-density electron beam that is compressed
by a strong magnetic field produced by magnetic coils. An electron gun with a cathode of high emissivity produces the electron beam. The ions that are produced by successive electron impact ionizations inside the electron beam are radially confined by the negative space charge of the electron beam, and the axial confinement of the ions is achieved by superposition of electrostatic potentials generated using a minimum of three collinear drift tubes. The higher potentials on the outer drift tubes create an axial electrostatic trap. Hence, the ionization time of the ions can be controlled by switching the axial trap potentials periodically on and off.

The magnetic field for the electron beam compression is generated by a special magnet system, which can consist of superconducting magnetic coils or permanent magnet rings. The electron beam is dumped in an electron beam collector where the electrons are separated from the extracted ions. Usually EBISs are equipped with one or more radial ports. This allows for spectrometry of electromagnetic radiation produced in the region of the ion trap during the ionization process.

\begin{figure}[htbp!]
\centering\includegraphics[width=0.6\linewidth]{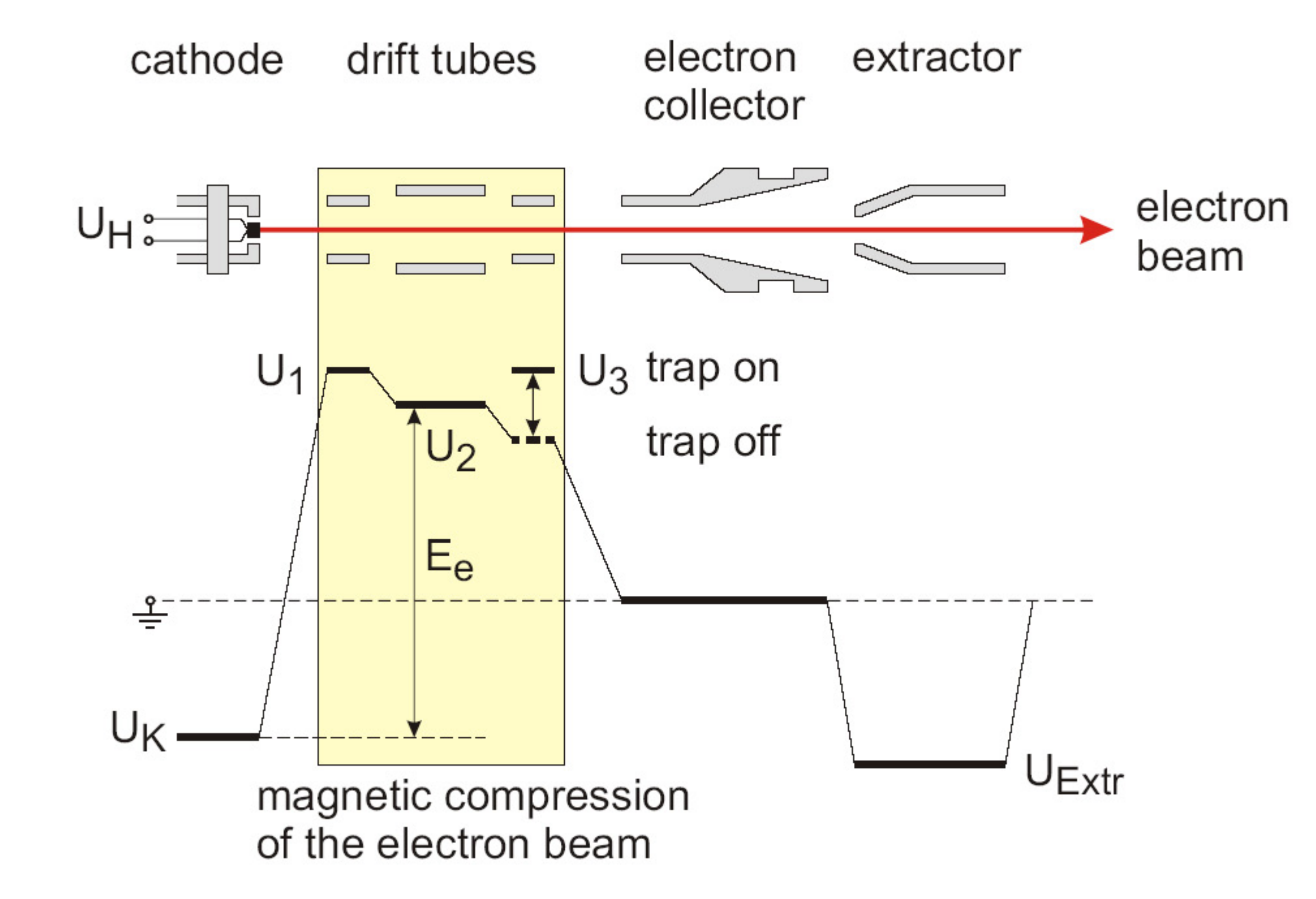}
\caption{Electrical scheme of EBIS operation: $U_{\rm K}$, cathode potential; $U_{\rm Extr}$, ion extraction potential; $U_{\rm H}$, potential at the cathode heater; $U_1$, trap potential at the side of the electron gun; $U_3$, trap potential at the ion extraction side (switched for pulsed ion extraction); $U_2$, potential at the middle drift tube.}
\label{fig:elektrik1}
\end{figure}

The electrical scheme of EBIS operation is shown in Fig.~\ref{fig:elektrik1} for an EBIS with three drift tubes (solutions are also known with more drift tubes). The electron energy $E_\rme$ in the centre of the drift tubes where the ions are produced and accumulated is
\begin{equation}\label{lab:g1}
E_\rme = e ( U_2 - U_{\rm K} + U_\rme ),
\end{equation}
where $U_\rme$ is the space-charge potential of the electron beam, which slightly decreases the final electron beam energy.
The axial ion trap is controlled by switching the potential $U_3$ in the indicated
manner. The chosen dynamics and  height of the potential wall formed by the voltage $U_3$ allows the following three principal modes of source operation.
\begin{itemize}
\item  \textit{Permanently opened trap -- transmission mode.}
The trap is permanently open and ions are produced in the electron beam without axial trapping. This mode delivers high currents of the lowest charged ions with ion beam currents up to microamps.
\item \textit{Partially closed trap -- leaky mode.}
Selecting a low axial potential wall, a certain number of ions with adequate kinetic energy can overcome the potential wall and are extracted continuously. This mode delivers ions with preferentially low up to intermediate ion charge states and a small fraction of higher ion charge states, with ion beam currents up to nanoamps.
\item \textit{Periodically opened and closed trap -- pulsed mode.}
In this case the voltage $U_3$ is high enough to trap all ions axially. Periodic opening of the trap releases pulses of ions extracted with typical pulse widths in the order of some microseconds and allows the production of the highest currents of HCIs (up to microamps during the pulse).
\end{itemize}
Table \ref{tab:comp} compares important practically realized parameters of cryogenic and room-temperature EBISs.

\begin{table}[htb]
\caption{Parameters of cryogenic and room-temperature EBISs. \label{tab:comp}} \centering
\begin{tabular}{p{4cm}p{4.5cm}p{5.8cm}}
 \hline\hline
 Parameter & Cryogenic EBIS & Room-temperature EBIS \\
 \hline
 Magnetic system & superconducting coils\newline (3--8{\u}T) on axis & permanent magnets (SmCo, NdFeB)\newline  (250--600{\u}mT) on axis\\[1ex]
 Electron beam currents & up to amps & $\le$200{\u}mA\\[1ex]
 Electron beam energies & up to 200{\u}keV & up to 30{\u}keV\\[1ex]
 Electron beam densities & $j_\rme > 1000${\u}A{\u}cm$^{-2}$ & $j_\rme < 500${\u}A{\u}cm$^{-2}$\\[1ex]
 Basic vacuum  & up to $10^{-12}${\u}mbar & up to $10^{-10}${\u}mbar\\[1ex]
 Length of the \newline ionization region & up to metres & $\le$6{\u}cm
 \newline \\[1ex]
 Ion charge states & highest ion charge states \newline Xe$^{54+}$, U$^{92+}$ at maximum   &  bare ions up to $Z = 28$, \newline Kr$^{34+}$, Xe$^{48+}$, Au$^{60+}$, etc.\\[1ex]
 Set-up time & $\ge$1{\u}day & hours\\[1ex]
 Remarks & large devices \newline liquid helium cooling \newline latest developments: \newline \mbox{\quad}refrigerator cooling & compact \newline transportable \newline low initial and maintenance costs \newline short set-up times  \\[1ex]
 \hline\hline
\end{tabular}
\end{table}

\subsection{The history of EBISs: a short retrospect}

EBIT/EBIS ion sources have now been available as reliable sources of HCIs for about 20 years. After the first demonstration of the working principle of an electron beam ion source (EBIS) in the pioneer works of Donets, Ovsyannikov and collaborators \cite{bib:Donets1967,bib:Donets1979,bib:Donets1981}, the development of the first compact electron beam ion trap (EBIT) was reported by Levine \textit{et al.} \cite{bib:Levine1988}. Starting from that date, many different EBIT/EBIS ion sources have been developed and put into operation at different places all over the world.

First of all, it seems to be necessary to clear up the terms `EBIT' and `EBIS'. Both terms can be found in the literature. The term EBIT stands for electron beam ion trap. Since all EBITs can in principle also work like an EBIS, it seems to be more common to speak about an EBIS unless the function as an ion trap is of interest, as is the case in spectroscopic studies on HCIs.

Making no claim to be complete, sources such as the LLNL EBIT, the NIST EBIT, the Tokyo EBIT, the Shanghai EBIT and the Heidelberg EBIT with their different modifications should be mentioned here. Furthermore, the first attempts to built a refrigerated (cryogenic closed cycle, no liquid helium) EBIT
with a 2{\u}cm ion trap using outer correction coils are known \cite{bib:McDonald2005} and were realized with the Stockholm EBIT. Most of these EBITs are special constructions realized by several laboratories with the
goal to perform specific investigations on highly charged ions.

Another line of development is the series of compact room-temperature
EBIT/EBIS ion sources of the Dresden EBIT/EBIS family \cite{bib:Ovsyannikov1999,bib:dreebit}. They have a patented
operation principle that allows the production of highly charged ions efficiently and with ion beam currents competitive with other known EBIT/EBIS solutions. Continuing this way a new refrigerated superconducting
ion source, the Dresden EBIS-SC, was constructed and commissioned in 2010 \cite{bib:Zschornack2010}. In this context
the basic idea was to build an ion source working with a liquid-helium-free refrigerator technique for cooling the superconductors that generate the strong magnetic field compressing the electron beam and to realize a very compact design of the source body. In spite of the compact ion source assembly,
the system has proved successful for source operation with electron beams up to 700{\u}mA.

An overview of EBIS installations in 2010 and earlier can be found in Ref.~\cite{bib:Becker2010}. Nevertheless, the number of employed EBIS/EBIT is increasing permanently. In the last two years installations in Kielce and Krakow (both Poland), Huddersfield (Great Britain), Clemson (USA), Dresden-Rossendorf and Jena (both Germany) and Shanghai (China) have emerged.

Most of the EBISs of international reputation are special laboratory constructions, where each EBIS has its own particularities. Commercial versions of EBIT and EBIS devices are known only from two vendors:
\begin{itemize}
\item \textit{Physics \& Technology (Livermore, USA).}
Offered is a so-called REBIT/S (Refrigerated Electron Beam Ion Trap/Source; for details see \cite{bib:McDonald2005}).
\item \textit{Dreebit GmbH (Dresden, Germany).}
Offered is a series of compact and efficient room-temperature EBIT and EBIS sources (Dresden EBIT, Dresden EBIS and Dresden EBIS-A) as well as a superconducting EBIS (Dresden EBIS-SC), using refrigerating technology for magnet cooling (for details see \cite{bib:dreebit}).
\end{itemize}

\section{EBISs: the basic physics}

\subsection{The ionization balance}

The intended purpose of an EBIS is to produce highly charged ions. To produce HCIs a high-density electron beam acts as ionization medium. For neutrals and ions in the electron beam, the balance between
\begin{itemize}
\item charge generating processes and
\item charge destructive processes
\end{itemize}
is of importance for reaching a certain ionization stage.

For ion production the dominant process is electron impact ionization.
Therefore, only this process is considered at first. For the ion production in an EBIS we can write
\begin{equation}\label{g24-1}
\begin{array}{llll}
\mbox{to $q = 1$} \qquad\quad & \displaystyle\frac{\rmd n_0}{\rmd t}
& = & -\lambda_1 n_0 , \\[2ex]
~~\vdots & \quad\vdots & & \qquad\vdots\\
\mbox{to $q$} & \displaystyle\frac{\rmd n_q}{\rmd t}
& = & \underbrace{\lambda_q n_{q-1}}_{\mbox{\footnotesize ion production}}
    - \underbrace{\lambda_{q+1} n_q}_{\mbox{\footnotesize ion destruction}} , \\[2ex]
~~\vdots & \quad\vdots & & \qquad\vdots\\
\mbox{to $Z$} & \displaystyle\frac{\rmd n_Z}{\rmd t}
& = & \lambda_Z n_{Z-1} ,
\end{array}
\end{equation}
where $n_0$ is the neutral particle density. For $\lambda_q$ we have that
\begin{equation}\label{g24-2}
 \lambda_q = \sigma_q v_\rme n_\rme \qquad\quad \mbox{(reaction rate)},
\end{equation}
where $\lambda_q$ has the dimension of s$^{-1}$ and $\sigma_q$ is the ionization cross-section, $v_\rme$ the electron velocity and $n_\rme$ the electron density.

If the ionization process starts only with neutral particles it yields
\[
t=0, \qquad\quad n_0(0)=n_0^0, \qquad\quad n_q(0)=0.
\]
This leads to the solution
\begin{eqnarray}\label{g24-3}
n_0 & = & n_0^0 \, \rme^{-\lambda_1\,t} , \nonumber \\[1ex]
n_1 & = & \displaystyle\frac{n_0^0 \lambda_1}{\lambda_2-\lambda_1}
            (\rme^{-\lambda_1 t} - \rme^{-\lambda_2 t}) , \\[1ex]
n_2 & = & n_0^0 \lambda_1 \lambda_2
    \left(\displaystyle\frac{\rme^{-\lambda_1 t}} {(\lambda_2-\lambda_1)(\lambda_3-\lambda_1)}
    + \displaystyle\frac{\rme^{-\lambda_2 t}} {(\lambda_3-\lambda_2)(\lambda_1-\lambda_2)}
    + \displaystyle\frac{\rme^{-\lambda_3 t}} {(\lambda_1-\lambda_3)(\lambda_2-\lambda_3)}\right) , \nonumber\\
 \vdots & & \vdots\nonumber \\
n_q & = & n_0^0 \prod\limits_{l=1}^q \lambda_l \left(\sum\limits_{j=1}^{q+1} \displaystyle\frac{\rme^{-\lambda_j t}}{\prod\nolimits_{k=1,\,k\neq q}^{q+1} (\lambda_k-\lambda_j)}\right) . \nonumber
\end{eqnarray}

In principle, Eq.~(\ref{g24-1}) can also be integrated for other conditions, such as, for example, for $n_q(0) \neq 0$, for charge breeding processes or for external ion injection into an EBIS.
With Eq.~(\ref{g24-2}) the production rate of ions with individual ion charge states can be determined.

If we develop the exponents in Eq.~(\ref{g24-3}) it yields
\begin{eqnarray}\label{g24-4}
n_0 & = & n_0^0 (1-\lambda_1 t) , \nonumber \\[1ex]
n_1 & = & n_0^0 \lambda_1 t , \nonumber \\[1ex]
n_2 & = & n_0^0 \lambda_1 \lambda_2 \displaystyle\frac{t^2}{2} , \\[1ex]
\vdots & & \vdots\nonumber\\
n_q & = & n_0^0 \displaystyle\frac{t^q}{q!} \prod\limits_{j=1}^q \lambda_j . \nonumber
\end{eqnarray}
The characteristic time $\tau$ for ion production (ionization time) follows using
\begin{equation}\label{g24-5}
t_q = \displaystyle\frac{1}{\lambda_q} = \frac{1}{\sigma_q v_\rme n_\rme} \qquad\quad \left(\frac{1}{\mbox{reaction rate}}\right) ,
\end{equation}
to give
\begin{eqnarray}\label{g24-6}
\tau_1 & = & t_1 , \nonumber \\[1ex]
\tau_2 & = & t_1+t_2 , \\[1ex]
\vdots & & \vdots\nonumber\\
\tau_q & = & \sum\limits_{k=1}^q t_q . \nonumber
\end{eqnarray}

For estimating when a certain ionization stage reaches its maximum in the actual ion charge-state distribution, Eq.~(\ref{g24-6}) can be used.
Hence we write
\begin{equation}\label{g24-7}
j_\rme \tau_q = \sum\limits_{k=1}^q \displaystyle\frac{e}{\sigma_k} \qquad\quad (\mbox{ionization factor}),
\end{equation}
with
\[
j_\rme = n_\rme v_\rme
\]
as the electron beam density (A{\u}cm$^{-2}$).
The ionization factor is a fundamental quantity that determines the ionization stage in an ion source.

The production of a mean ion charge state $\bar{q}$ is only possible if a certain $j \tau$ value (ionization factor) is reached. The following must hold:
\begin{equation}\label{g24-8}
j_\rme \tau_q \ge \sum\limits_{k=1}^q \displaystyle\frac{e}{\sigma_k}.
\end{equation}
The precondition must be an electron energy $E_\rme$ higher than the ionization potential $I_q$ (ideally about (2--3)${\times}I_q$).

\begin{figure}[htbp!]
\centering\includegraphics[width=0.7\linewidth]{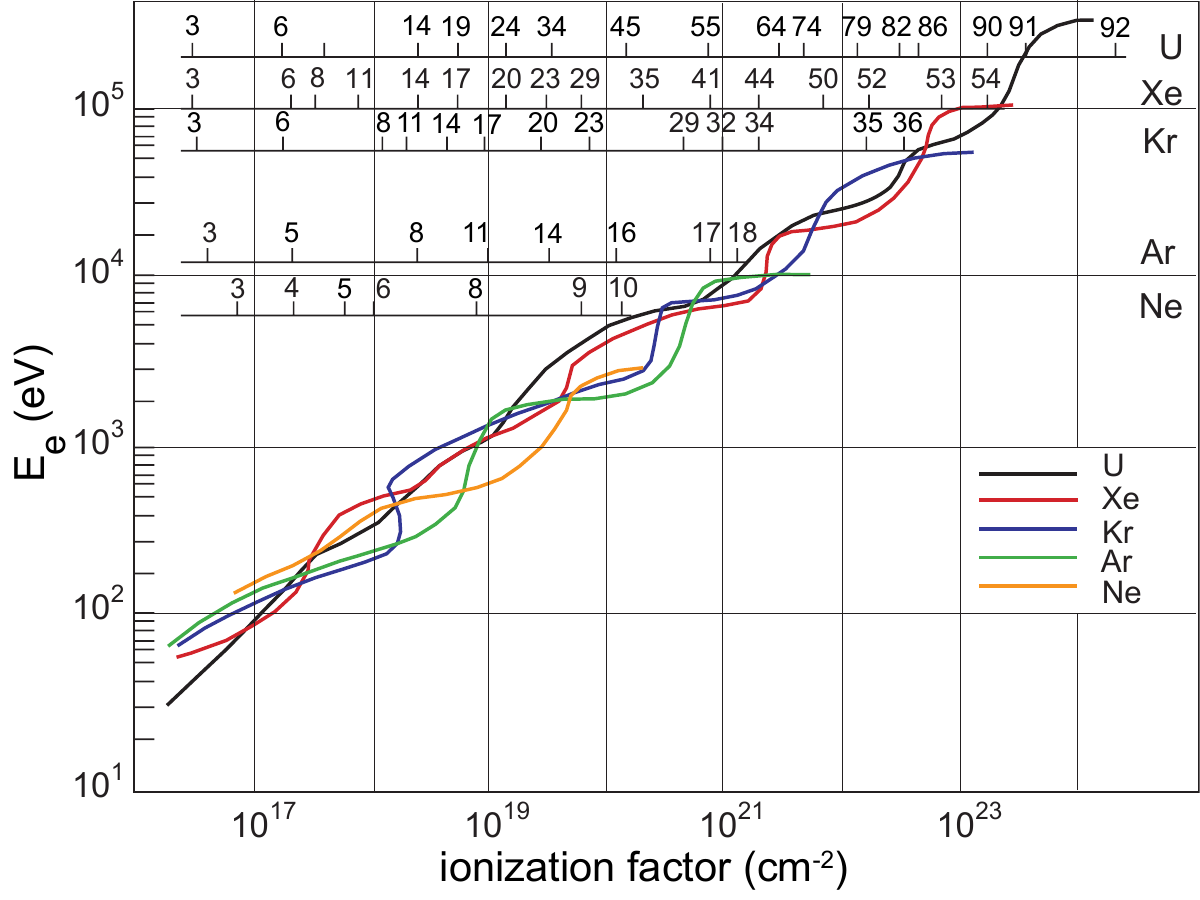}
\caption{Interrelation between ionization factor and electron energy for the effective production of ions in different ion charge states (according to an idea of Prof.\ E.~D.\ Donets (Dubna)).} 
\label{b24-1}
\end{figure}

The best values for the ionization factor and the electron beam energy under consideration of the excitation function of the electron-impact ionization cross-sections for selected elements are shown in Fig.~\ref{b24-1}.
Thus producing high ionization stages requires
\begin{itemize}
\item a sufficient electron beam energy and
\item a sufficiently high ionization factor.
\end{itemize}

\begin{figure}[htbp!]
\centering\includegraphics[width=0.65\linewidth]{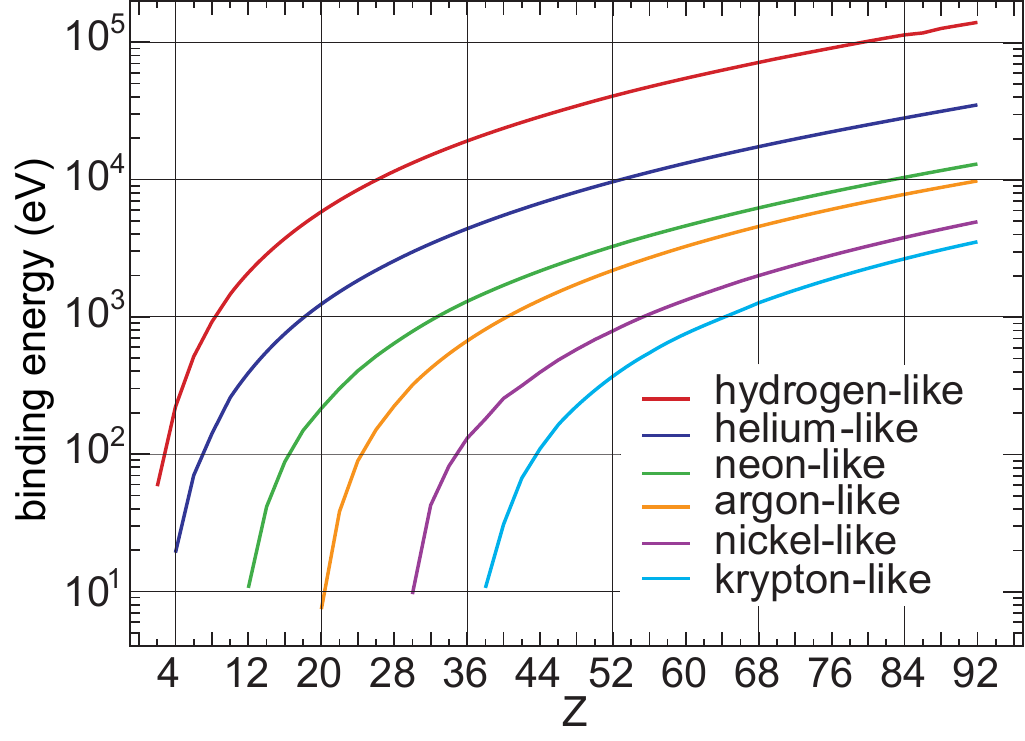}
\caption{Electron binding energies for the ion production of different isoelectronic sequences} 
\label{jtau-pse}
\end{figure}

In order to explain the role of electron binding energies, in Fig.~\ref{jtau-pse} we show in more detail the electron binding energies of ions for different isoelectronic sequences as a function of the atomic number. Thereby it is considered that the ionization cross-section has its maximum at about 2.7~times the electron binding energy of the electron to be ionized.

In practice, processes such as
\begin{itemize}
\item charge exchange and
\item ion losses from the source
\end{itemize}
can reduce the ionization power significantly.
Of special importance in this case is the charge exchange with neutrals.
The ionization factor as a function of the ionization time and the electron energy for selected isoelectronic sequences is given in Table ~\ref{ionisation-a} (according to Ref.~\cite{bib:Ovsyannikov1996}).

\begin{figure}[htbp!]
\centering
\centering\includegraphics[width=0.70\linewidth]{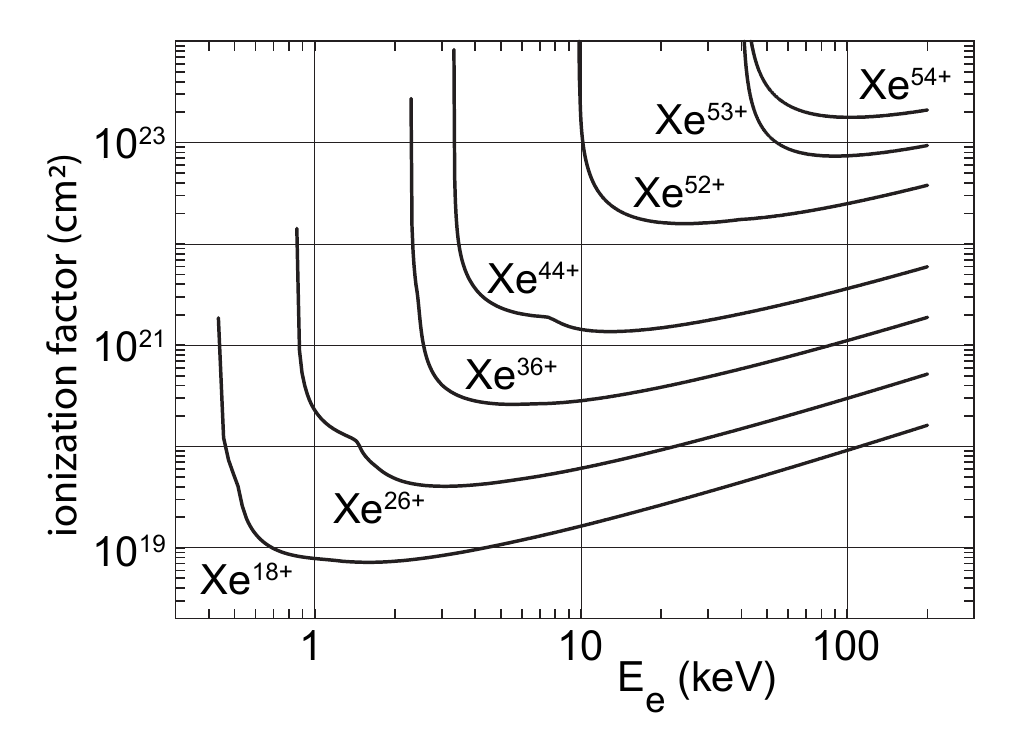}
\caption{Ionization factors for the production of individual xenon ionization charge states}
\label{jtau-xe}
\end{figure}

Figure \ref{jtau-xe} shows this behaviour for xenon ions. The ionization factor for the production of a certain ion charge state is minimal at 2.7~times the electron binding energy and increases again at higher electron energies. To produce the highest ionization charge states, ionization factors many orders of magnitude higher than those ones for low charged ions must be reached. It should be noted again that the issues discussed are only valid under the idealized condition that only electron impact ionization takes place.

\begin{table}[thb]
\centering
\caption{Ideal production conditions for ions of different isoelectronic sequences. Given are the ionization factor $j_\rme \tau$ (e$^-${\u}cm$^{-2}$), the optimal electron beam energy (keV) and the required ionization time (ms or s) for an assumed ionization factor of $j_\rme \tau  = 3\times 10^{22}${\u}e$^-${\u}cm$^{-2}$. \label{ionisation-a}}
\begin{tabular}{p{1.9cm}p{1.7cm}p{1.7cm}p{1.7cm}p{1.7cm}p{1.7cm}p{1.7cm}}
 \hline\hline
 Sequence & Neon\newline $Z=10$ & Argon\newline $Z=18$& Krypton\newline $Z=36$
 & Xenon\newline $Z=54$& Gold\newline $Z=79$& Uranium\newline $Z=92$\\
 \hline
 \noalign{\vspace{4pt}}
 Atom\newline fully\newline ionized & Ne$^{10+}$\newline $2 \times 10^{21}$\newline 3   \newline 7 ms&
 Ar$^{18+}$ \newline $2 \times 10^{21}$ \newline 9  \newline 67 ms&
 Kr$^{36+}$\newline $3 \times 10^{22}$\newline 40   \newline 1 s&
 Xe$^{54+}$\newline $2 \times 10^{23}$\newline 80  \newline 7 s&
 Au$^{79+}$ \newline $6 \times 10^{23}$\newline 180   \newline 20 s&
 U$^{92+}$\newline $2 \times 10^{24+}$\newline 300  \newline 67 s\\[12pt]
 Helium-\newline like&
 Ne$^{8+}$\newline $8 \times 10^{18}$\newline 0.6  \newline 0.3 ms&
 Ar$^{16+}$\newline $ 1 \times 10^{20}$\newline 2   \newline 3 ms &
 Kr$^{34+}$ \newline $ 2 \times 10^{21}$ \newline 7   \newline 67 ms&
 Xe$^{52+}$\newline $2 \times 10^{22}$\newline 20  \newline 0.7 s&
 Au$^{77+}$\newline $6 \times 10^{22}$ \newline 45  \newline 2 s&
 U$^{90+}$\newline $2 \times 10^{23}$\newline 70   \newline 7 s\\[12pt]
 Neon-\newline like & &
 Ar$^{8+}$ \newline $3 \times 10^{18}$ \newline 0.3  \newline 0.1 ms&
 Kr$^{28+}$ \newline $3 \times 10^{20}$\newline 4   \newline 10 ms&
 Xe$^{44+}$\newline $2 \times 10^{21}$ \newline 8   \newline 67 ms&
 Au$^{69+}$ \newline $6 \times 10^{21}$ \newline 17   \newline 200 ms&
 U$^{82+}$ \newline $3 \times 10^{22}$ \newline 30   \newline 1 s\\[12pt]
 Argon-\newline like & & &
 Kr$^{18+}$ \newline $1 \times 10^{19}$\newline 1.5   \newline 0.3 ms&
 Xe$^{36+}$\newline $2 \times 10^{20}$ \newline 5   \newline 7 ms&
 Au$^{61+}$ \newline $1 \times 10^{21}$ \newline 12   \newline 33 ms&
 U$^{74+}$ \newline $5 \times 10^{21}$ \newline 20   \newline 167 ms\\[12pt]
 Krypton-\newline like & & & &
 Xe$^{18+}$\newline $6 \times 10^{18}$ \newline 1   \newline 0.4 ms&
 Au$^{43+}$ \newline $1 \times 10^{20}$ \newline 4   \newline 3 ms&
 U$^{56+}$ \newline $7 \times 10^{20}$ \newline 7   \newline 23 ms\\[12pt]
 Xenon-\newpage like& & & & &
 Au$^{25+}$ \newline $2 \times 10^{19}$ \newline 1.5   \newline 0.7 ms&
 U$^{38+}$ \newline $7 \times 10^{19}$ \newline 4   \newline 2 ms\\[4pt]
 \hline\hline
\end{tabular}
\end{table}

\subsection{Atomic physics for ion production}\label{basisprozesse}

\subsubsection{Overview}

The following are the atomic processes contributing to ion production;
\begin{enumerate}
\item electron impact ionization,
\item charge exchange,
\item recombination,
\item three-body recombination,
\item dielectronic recombination,
\item photoionization,
\item vacancy cascades and
\item electron shake-off processes.
\end{enumerate}
Later on we will discuss these further processes that are important for ion losses and the energy balance in EBISs:
\begin{itemize}
\item ion heating by elastic electron collisions,
\item ion--ion energy exchange,
\item ion confinement and
\item ion losses from the trap.
\end{itemize}
In the following we will discuss the most important processes in more detail. For the basic atomic processes, some useful estimation formulas are given. Nevertheless, we note that these formulas do not completely substitute more precise quantum-mechanical calculations for individual processes.

\subsubsection{Direct Coulomb ionization}

\begin{figure}[htbp!]
\centering\includegraphics[width=0.75\linewidth]{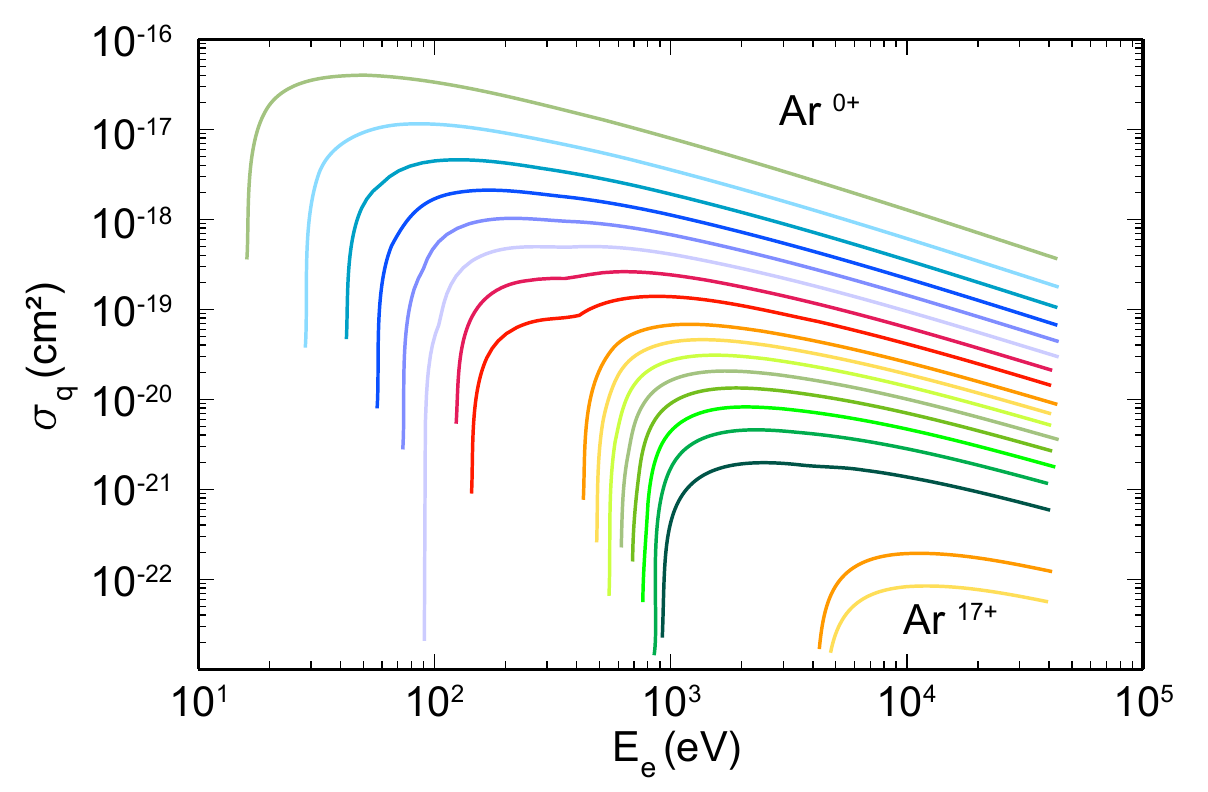}
\caption{Single electron impact ionization cross-sections as a function of the electron energy (according to Ref.~\cite{bib:Werner1997})
\label{b24-3}}
\end{figure}

In simple cases, the cross-sections for single electron impact ionization
\[ 
\mathrm{X}^{q+} + \rme^- \rightarrow  \mathrm{X}^{(q+1)+} + 2\rme^-
\]
can be estimated for different subshells $j$ by the Lotz formula \cite{bib:Lotz1967}
\begin{equation}\label{g24-16}
\sigma_{qj} = a_{qj} g_{qj} \displaystyle\frac{\ln(E_\rme/I_{qj})}{E_\rme I_{qj}} \left\{ 1 - b_{qj} \exp\left[ - c_{qj}\left(\displaystyle\frac{E_\rme}{I_{qj}} - 1\right)\right]\right\}.
\end{equation}
For all ionization processes, the electron energy $E_\rme$ must be greater than the related ionization potential $I_{qj}$, otherwise the ionization cross-section is equal to zero.
The quantities $a_{qj}$, $b_{qj}$ and $c_{qj}$ are tabulated constants. For highly charged ions, they are given by
\[ 
a_{qj} = 4.5 \times 10^{-14}{\u}\mbox{cm}^2{\u}\mbox{eV}^2
\]
and
\[ 
b_{qj} = c_{qj} = 0 ,
\]
with $g_{qj}$ as the occupation number of the subshell  $j$ of an ion in the charge state  $q$. The quantity
$I_{qj}$ is the ionization potential for the ionization of the subshell $j$.
The total ionization cross-section results as a sum over all occupied subshells:
\begin{equation}\label{g24-17}
\sigma_q(E_\rme) = \sum\limits_q \sigma_{qj}(E_\rme).
\end{equation}
As an example, the electron impact ionization cross-sections for argon ions are shown in Fig.~\ref{b24-3}~\cite{bib:Werner1997}.

We will not go into the details for the double ionization process
\[ 
\mathrm{X}^{q+} + \rme^- \rightarrow  \mathrm{X}^{(q+2)+} + 3\rme^-
\]
but we  refer the reader to the estimation formula from Mueller and Frodel \cite{bib:Mueller1980}, which was derived from experimental data. As a rule double ionization cross-sections for most species are about one order of magnitude or more lower than cross-sections for single ionization  cross-sections.

\subsubsection{Charge exchange}

The charge exchange between ions can be written as
\[ 
\mathrm{X}^{q+} + \mathrm{X}^{p+} \rightarrow \mathrm{X}^{i+} + \mathrm{X}^{(q+p-i)+} 
\]
and correspondingly for the charge exchange with a neutral atom
\[ 
\mathrm{X}^{q+} + \mathrm{X} \rightarrow \mathrm{X}^{(q-1)+} + \mathrm{X}^+ 
\]
Owing to the fact that
\[ 
\mbox{X}^+ + \mbox{X} \rightarrow \mbox{X} + \mbox{X}^+ 
\]
charge exchange does not play a role for singly charged ions.

\begin{figure}[htbp!]
\centering
\centering\includegraphics[width=0.7\linewidth]{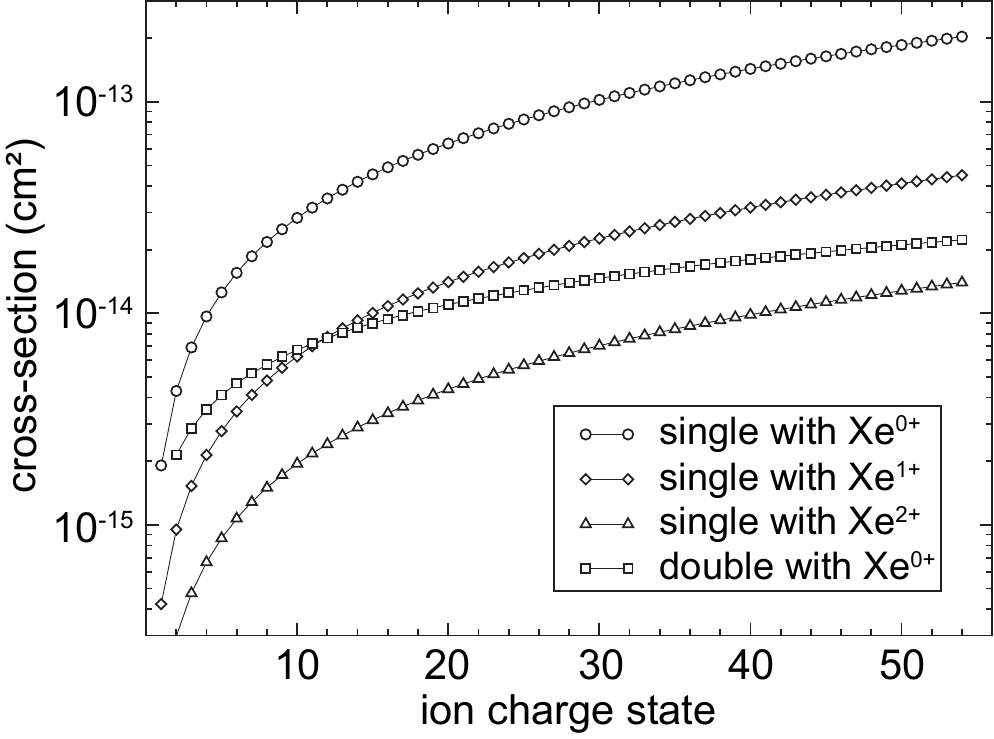}
\caption{Charge exchange cross-sections for single charge exchange with Xe$^{(0,1,2)}$ ions as well as for the double charge exchange with xenon neutrals (after Ref.~\cite{bib:Ullmann2005}).}
\label{b24-5}
\end{figure}

Of special importance in this case is the charge exchange of ions with neutrals due to the very high cross-sections. This circumstance is important for EBIS operation and can limit the reachable ionization stages in the electron beam.
For collision energies below 25{\u}keV/u, charge exchange can be estimated by the formula from Mueller and Salzborn \cite{bib:Mueller1976}
\begin{equation}\label{g24-20}
\sigma_{q\,\rightarrow\,q-1}\ \mbox{(cm$^2$)} \approx (1.43 \pm 0.76) \times 10^{-12} \, q^{1.17} \, [I_q\,\mbox{(eV)}]^{-2.76}
\end{equation}
and is of the order of  $10^{-14}\mbox{--}10^{-15}${\u}cm$^{-2}$. As a rule, charge exchange is the dominant loss mechanism for highly charged ions and is especially important if the gas pressure in the ion source is relatively high. As we can see from Eq.~\eqref{g24-20}, the cross-section is independent of the energy of the projectiles.
In Fig.~\ref{b24-5} cross-sections for single and double charge exchange are given for xenon ions.

\subsubsection{Radiative recombination}

\begin{figure}[htbp!]
\centering
\centering\includegraphics[width=0.2\linewidth]{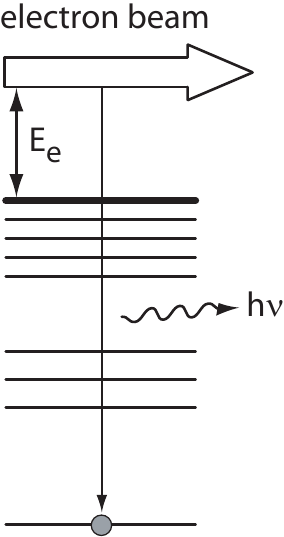}
\caption{Scheme of the radiative recombination process. The energy  $\hbar \omega$ of the emitted photon is calculated as $\hbar \omega = E_\rme - E_{\rm B}$, where $E_{\rm B}$ is the electron binding energy.}
\label{b24-6}
\end{figure}

The scheme for radiative recombination (RR)
\[  
\mathrm{X}^{q+} + \rme^- \rightarrow  \mathrm{X}^{(q-1)+} + \hbar \omega 
\]
is shown in Fig.~\ref{b24-6}.

\begin{figure}[htbp!]
\centering\includegraphics[width=0.75\linewidth]{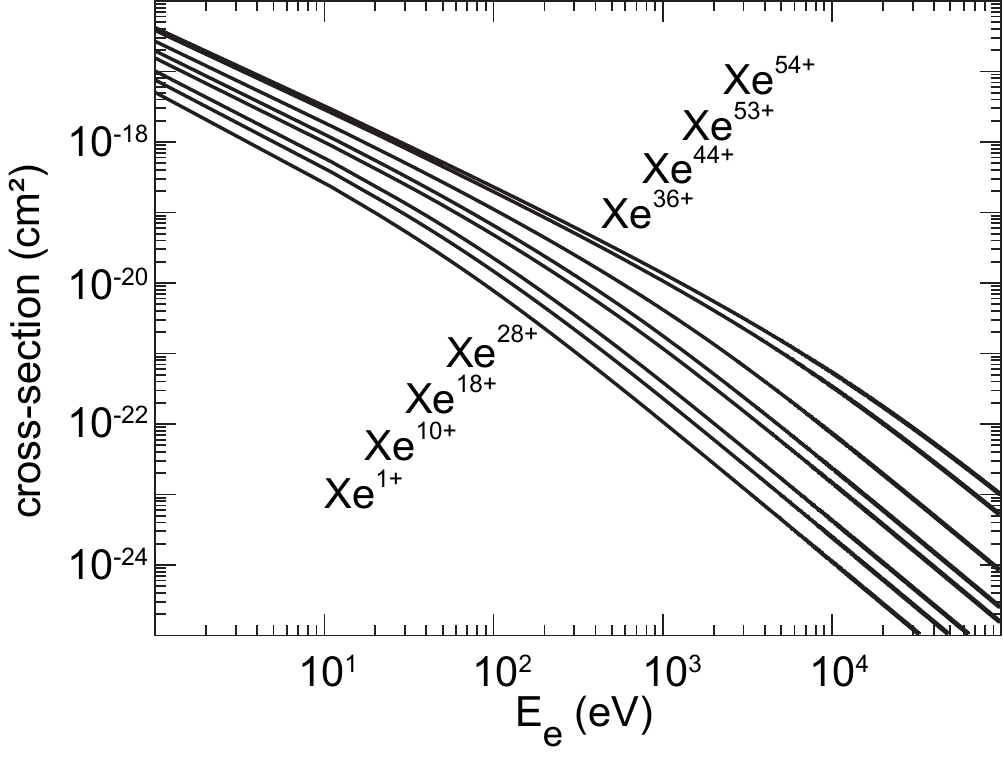}
\caption{Cross-sections for the radiative recombination on xenon ions (according to Ref.~\cite{bib:Ullmann2005})}
\label{b24-7}
\end{figure}

The cross-section for charge exchange processes can be estimated by the formula from Kim and Pratt \cite{bib:Kim1983}
\begin{equation}\label{g24-26}
\sigma_q^{\rm RR}(E_\rme) = \displaystyle\frac{8 \pi}{3 \sqrt{3}} \alpha \lambda_\rme^2 \chi_q (E_\rme) \ln\left(1+\displaystyle\frac{\chi_q(E_\rme)}{2\hat{n}}\right),
\end{equation}
with
\[ 
\chi_q(E_\rme) = (Z+q)^2 \frac{I_{\rm H}}{4 E_\rme}
\]
and
\[ 
\hat{n} = n + (1-W_n) - 0.3.
\]
Here $n$ is the main quantum number of the valence shell, $W_n$ is ratio of the number of unoccupied states to the total number of states in the valence shell, $\lambda_\rme = 3.861 \times 10^{-11}${\u}cm (electronic Compton wavelength) and $I_{\rm H}$ = 13.6{\u}eV.
As an example we show the cross-sections for radiative recombination on xenon ions as a function of the electron energy in Fig.~\ref{b24-7}.

Since RR processes counteract the ionization for each element and ion charge state, there exists a characteristic electron energy at which the ionization rate is equal to the RR rate. This behaviour is discussed in more detail in Ref.~\cite{bib:Becker2009}.

\subsubsection{Further processes}

Besides the main processes discussed above, in special cases some further contributions from
\begin{itemize}
\item dielectronic recombination (DR),
\item three-body recombination,
\item vacancy cascades and
\item photoionization
\end{itemize}
are of interest, but they will not be discussed here.

\subsection{Balance equation for the ion charge-state distribution}

The number of ions with charge  $q$ produced from ions with charge  $(q-1)$ and ionized further to the charge state  $(q+1)$ can be described by a balance equation  for successive electron impact ionization and competing processes:
\begin{eqnarray}
\label{g26-42}
\displaystyle\frac{\rmd n_q}{\rmd t} & = & R_{q-1\,\rightarrow\,q}^{\rm ion} - R_{q\,\rightarrow\,q+1}^{\rm ion} + R_{q+1\,\rightarrow\,q}^{\rm RR} - R_{q\,\rightarrow\,q-1}^{\rm RR} \nonumber \\
 & & {} + R_{q+1\,\rightarrow\,q}^{\rm exch}  - R_{q\,\rightarrow\,q-1}^{\rm exch} - R_{q}^{\rm ax, esc} - R_q^{\rm rad, esc} + R^{\rm source},
\end{eqnarray}
with ionization terms $R^{\rm ion}$,
radiative recombination terms $R^{\rm RR}$,
charge exchange terms $R^{\rm exch}$,
axial ion loss terms $R^{\rm ax, esc}$,
radial ion loss terms $R^{\rm rad, esc}$ and
particle source term $R^{\rm source}$.

If the ionization process occurs in single-step processes and if we neglect the axial loss (which is a good approximation for appropriate field configurations) this yields
\begin{eqnarray}\label{g26-43}
\frac{\rmd n_0}{\rmd t} & = & \underbrace{-n_0 \lambda_{0,1}}_{\mbox{\footnotesize ionization}} + \underbrace{n_1 \lambda_{1,0}}_{\mbox{\footnotesize charge exchange}} , \nonumber\\
\frac{\rmd n_1}{\rmd t} & = & n_0 \lambda_{0,1} - n_1 (\lambda_{1,2} + \lambda_{1,0}) + n_2 \lambda_{2,1} - \left[\frac{\rmd n_1}{\rmd t}\right]^{\rm rad, esc} , \nonumber\\
& \vdots & \\
\frac{\rmd n_q}{\rmd t} & = & n_{q-1} \lambda_{q-1,q} - n_q (\lambda_{q,q+1} + \lambda_{q,q-1} ) + n_{q+1} \lambda_{q+1,q} - \left[\frac{\rmd n_q}{\rmd t}\right]^{\rm rad, esc} , \nonumber\\
\frac{\rmd n_z}{\rmd t} & = & n_{Z-1} \lambda_{Z-1,Z} - n_Z \lambda_{Z,Z-1} - \left[\frac{\rmd n_Z}{\rmd t}\right]^{\rm rad, esc} . \nonumber
\end{eqnarray}
Here $n_0,\dots,n_Z$  are the densities for neutrals and ions of the charge state $q$. Further we have the ionization coefficients $\lambda_{q,q+1}$ and the recombination and charge exchange coefficients $\lambda_{q,q-1}$.
Details about the corresponding terms, the influence on the development of the individual ion densities, the development of the energy densities and ion evaporation of ions from the ionization volume can be found, for example, in Refs.~\cite{bib:Penetrante1991} and~\cite{bib:Kalagin1998}.

\begin{figure}[hbt]
\centering
\centering\includegraphics[width=0.75\linewidth]{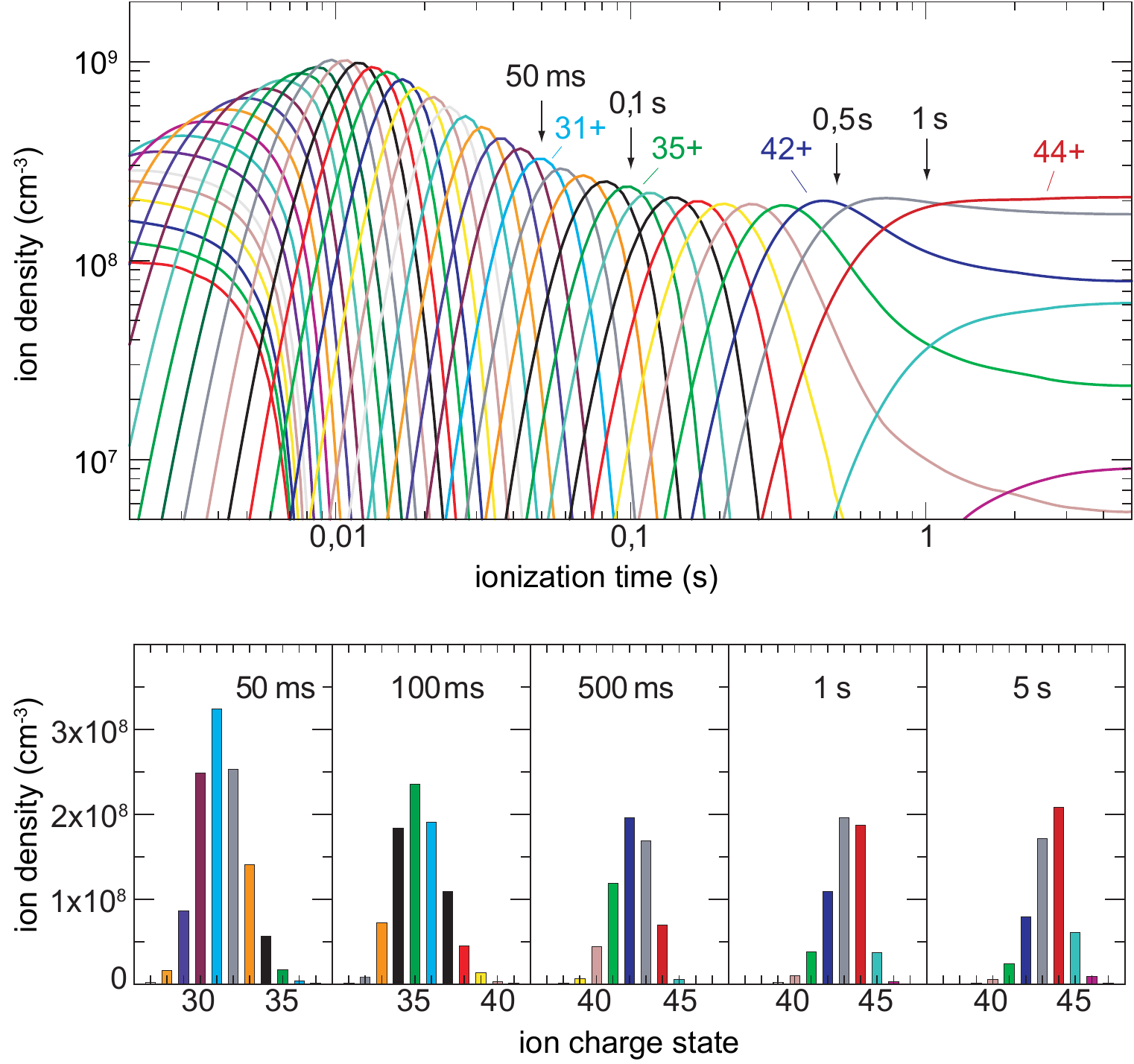}
\caption{Evolution of the ionization of xenon ions in a Dresden EBIT at $E_\rme = 15${\u}keV, $I_\rme = 40${\u}mA and $p = 2\times 10^{-9}${\u}mbar.}
\label{b26-4}
\end{figure}

In Fig.~\ref{b26-4}, as a typical example we show the evolution of xenon ion charge states in the electron beam of a Dresden EBIT as it follows from equation system~(\ref{g26-42}).

\subsection{Ion source physics: selected topics}

Owing to the restricted scope of this chapter, only some information about the important (but not complete) details of EBISs are given. For more detailed studies, please refer to the corresponding literature. For reasons of simplicity and copyright, in the following the author will give examples of results from EBISs of the Dresden EBIS type.

\subsubsection{Electrical trap capacity}

\begin{figure}[htbp!]
\centering\includegraphics[width=0.8\linewidth]{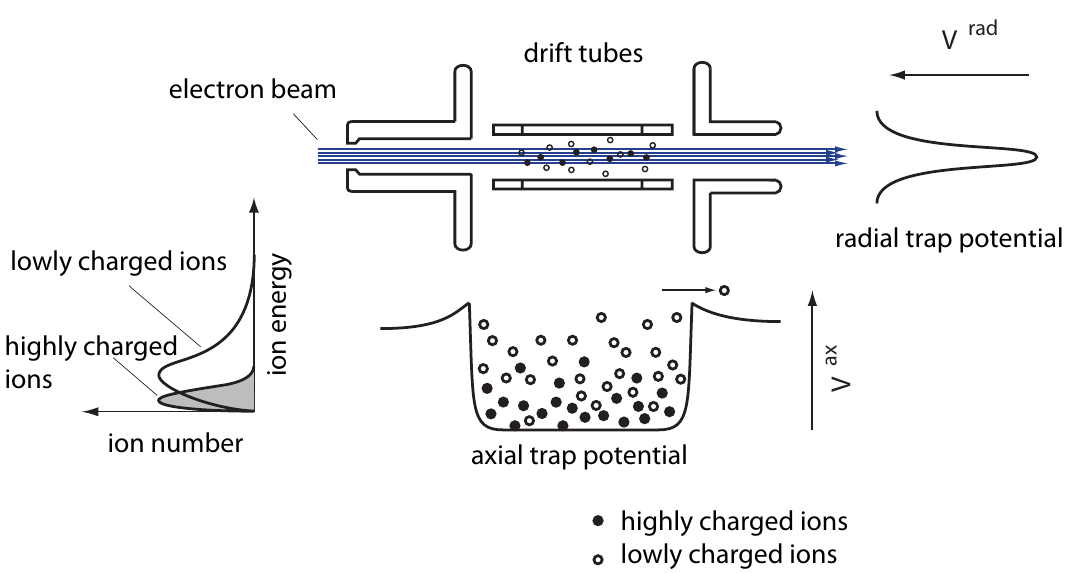}
\caption{Axial and radial potentials in an EBIT. Owing to the effective trap height according to $V_{\rm b} \times q$, highly charged ions are trapped more strongly than low charged ones. The energy exchange between the ions leads to an equilibrium distribution of the ion energy.}
\label{b26-1}
\end{figure}

The ion trapping in an EBIS (see Fig.~\ref{b26-1}) occurs
\begin{itemize}
\item in the axial direction by electrostatic potentials and
\item in the radial direction by the space-charge potential of the electron beam.
\end{itemize}

The total number of ions that can be stored in an EBIS is limited by the electrical trap capacity $C_{\rm el}$. For an estimation of $C_{\rm el}$, we assume a homogeneous electron beam that passes through an ion trap of  length  $L$ with the electron beam current $I_\rme$ and electron energy  $E_\rme$. Then the number of negative charges in the beam volume is a measure for the trap capacity $C_{\rm el}$. With
\[ 
I_\rme = \frac{\rmd Q}{\rmd t}, \qquad\quad v_\rme = \frac{\rmd x}{\rmd t}, 
\qquad\quad v_\rme = \sqrt{\displaystyle\frac{2E_\rme}{m_\rme}},
\]
we have
\[ 
\Delta Q = \frac{I_\rme \Delta x}{v_\rme} 
         = \frac{I_\rme L}{\sqrt{{2E_\rme}/{m_\rme}}}
\]
and for the trap capacity
\begin{equation}\label{g18-26c}
C_{\rm el} = 1.05 \times 10^{13} \, \displaystyle\frac{I_\rme \, (\mbox{A}) \, L \, (\mbox{m})}{\sqrt{E_\rme \, (\mbox{eV})}}.
\end{equation}
In order to estimate the useable trap capacity, in practice we must supplement Eq.~(\ref{g18-26c}) to
\begin{equation}\label{g18-26d}
C_{\rm el} = 1.05 \times 10^{13} \, \displaystyle\frac{I_\rme \, (\mbox{A}) \, L \, (\mbox{m})}{\sqrt{E_\rme \, (\mbox{eV})}} \, \alpha f ,
\end{equation}
with $\alpha$ the ratio  of ions with a certain ion charge state in the ion charge-state distribution and $f$ the charge compensation of the electron beam.

\subsubsection{Boltzmann distribution of the ion energy}

In the source volume ion--ion collisions lead very rapidly (in about milliseconds) to a  Boltzmann distribution of the ion energy
\begin{equation}\label{g26-19b}
f(E_\rmi) = \displaystyle\frac{2}{\sqrt{\pi}\ k T_\rmi} \sqrt{\displaystyle\frac{E_\rmi}{k T_\rmi}} 
\exp\left(-\displaystyle\frac{E_\rmi}{k T_\rmi}\right).
\end{equation}
The formation of a  Boltzmann energy distribution has consequences for the whole  ion trap process:
\begin{itemize}
\item \textit{Permanent ion losses.} At a certain mean ion energy, ions always exist with an energy greater than a critical energy
\begin{equation}\label{g26-20a}
E_{\rm cr} = q U_{\rm b},
\end{equation}
and can leave the trap with the barrier $U_{\rm b}$. This means that we have a constant ion loss from the trap. The ion loss ratio can be derived by integrating the distribution function. The behaviour of ions in the trap can be described by the non-stationary Boltzmann equation in a self-consistent model considering ionization and energy exchange processes.

\item \textit{Evaporative cooling} of multiply charged ions by light LCIs.
Elastic collisions between ions with different charge states and masses lead to an equilibrium energy distribution for each ion sort. Thereby the critical energy for leaving the ion trap for LCIs is lower than for highly charged ions, i.e.\ the ion lifetime in the ion trap is lower for LCIs. LCIs leave the beam first, i.e.\ they evaporate and transport the energy transmitted from the highly charged ions to lower charged ones out of the ionization volume. As a result of the process, the mean temperature of the highly charged ions decreases and ion losses of these ions are reduced.
\end{itemize}

\subsubsection{Radial trap potential}

\begin{figure}[h]
\centering\includegraphics[width=0.55\linewidth]{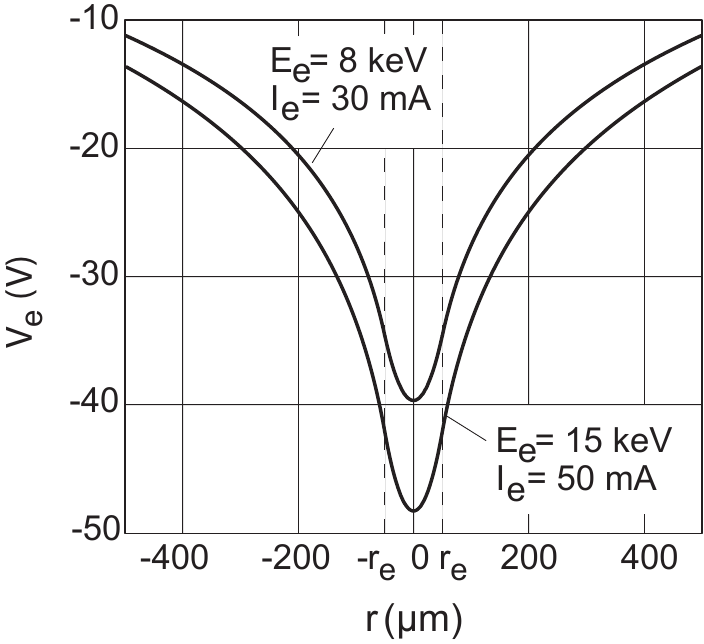}
\caption{Radial trap potential of a Dresden EBIT for two different electron beam currents}
\label{fall-pot}
\end{figure}

For simplicity, we consider an ion trap with a radial symmetric electron beam the electron density of which is homogeneously distributed over the radius $r_\rme$ of the electron beam.  Using Gauss's theorem the radial trap potential has the shape
\begin{equation}\label{g26-23}
V_\rme(r) =
\left\{
\begin{array}{ll}
\displaystyle\frac{U_\rme r^2}{r_\rme^2} & \mbox{for}\ r < r_\rme , \\[12pt]
U_\rme \left(2 \ln\displaystyle\frac{r}{r_\rme} + 1\right) & \mbox{for}\ r > r_\rme.
\end{array} \right.
\end{equation}
Here we have
\begin{equation}\label{g26-24}
 U_\rme = \displaystyle\frac{I_\rme}{4\pi\varepsilon_0 v_\rme} = \displaystyle\frac{1}{4\pi\varepsilon_0} \sqrt{\displaystyle\frac{m_\rme}{2}} \displaystyle\frac{I_\rme}{\sqrt{E_\rme}} .
\end{equation}
We can estimate $U_\rme$ with
\begin{equation}\label{g26-24a}
U_\rme \, (\mbox{V}) = \displaystyle\frac{30 I \, (\mbox{A})}{\sqrt{1 - \left( \displaystyle\frac{E_\rme \, (\mbox{keV})}{511} +1\right)^{-2}}}.
\end{equation}
As an example the  radial trap potential in Fig.~\ref{fall-pot} is shown for a typical operation regime of a Dresden EBIT.

The ion confinement is additionally influenced by the magnetic field enclosing the electron beam, so that we finally have an effective radial potential of
\begin{equation}\label{g26-22a}
 V_\rme^{\rm eff} (r) = V_\rme(r) + \displaystyle\frac{e q_\rmi B^2 r^2}{8 m_\rmi},
\end{equation}
with $R$ the drift tube radius.

\subsubsection{Ion heating and energy balance}

Ions  are heated in the electron beam by elastic Coulomb collisions and by microwave instabilities. If instabilities do not appear (huge Debye lengths), the main heating processes are elastic electron--ion collisions.

\paragraph{Elastic electron--ion collisions}

Ion heating by elastic electron collisions can be described by
\begin{equation}\label{g26-26}
\left[\displaystyle\frac{\rmd E_\rmi}{\rmd t}\right]^{\rm heat} =  (n_\rmi \sigma_\rmi v_\rme) \left(2 \displaystyle\frac{m_\rme}{m_\rmi} E_\rme\right) = 2   \displaystyle\frac{m_\rme}{m_\rmi} E_\rme f_{\rm ei},
\end{equation}
with
\[ 
t_{\rm ei} = f_{\rm ei}^{-1} 
= \frac{1}{n_\rmi \langle v_\rme \sigma_\rmi \rangle}
\]
and $\sigma_\rmi$ the Coulomb collision cross-section
\[ 
\sigma_\rmi = \frac{1}{(4\pi\varepsilon_0)^2} 4\pi \left(\frac{q_\rmi e^2}{m_\rme}\right)^2 \frac{\ln \Lambda_{\rm ei}}{v_\rme^4}.
\]
The Coulomb logarithm for electron--ion collisions  $\ln \Lambda_{\rm ei}$ has the shape
\begin{equation}\label{g26-27}
\ln \Lambda_{\rm ei} = \left\{
\begin{array}{ll}
23 - \ln \displaystyle\frac{q_\rmi \sqrt{n_\rme}}{(k T_\rme)^{3/2}} 
& \mbox{for}\  q_\rmi^2 \times 10{\u}\mbox{eV} > k T_\rme > \displaystyle\frac{m_\rme}{m_\rmi} k T_\rmi , \\[3ex]
24 - \ln \displaystyle\frac{\sqrt{n_\rme}}{(k T_\rme)} 
& \mbox{for}\  k T_\rme > q_\rmi^2 \times 10{\u}\mbox{eV} > \displaystyle\frac{m_\rme}{m_\rmi} k T_\rmi , \\[3ex]
30 - \ln \displaystyle\frac{q_\rmi^2 \sqrt{n_\rme}}{(k T_\rmi)^{3/2} A_\rmi} 
& \mbox{for}\  k T_\rme <  \displaystyle\frac{m_\rme}{m_\rmi} q_\rmi k T_\rmi,
\end{array}
\right.
\end{equation}
with $kT$ the energy (eV), $n$ the density (cm$^{-3}$), $A_\rmi$ the atomic mass and $k$ the Boltzmann constant. Energy transfer by Coulomb collisions then occurs with a rate
\begin{equation}\label{g26-28}
\left[\displaystyle\frac{\rmd(n_\rmi k T_\rmi)}{\rmd t} \right]^{\rm heat} \approx \left( -\displaystyle\frac{\rmd E_\rme}{\rmd t} \right) n_\rme .
\end{equation}

\paragraph{Elastic ion--ion collisions}
 
Energy transfer between ions occurs by elastic ion--ion collisions.  In these collisions energetic (`hot') ions are cooled by  colder ions. If we assume a  Boltzmann distribution of the ion energy, the cooling process can be characterized by
\begin{equation}\label{g26-29}
\left[\displaystyle\frac{\rmd E}{\rmd t} \right]^{\rm exch} = 2 \nu_{ij} n_i \displaystyle\frac{m_i}{m_j} \frac{k T_j - k T_i}{\left( 1 + \displaystyle\frac{m_i T_j}{m_j T_i}\right)^{3/2}}.
\end{equation}
Here the $\nu_{ij}$ are Coulomb collision rates between the ions
\begin{equation}\label{g26-30}
\nu_{ij} = \frac{1}{(4\pi\varepsilon_0)^2} \frac{4 \sqrt{2\pi}}{3} n_j \left(\frac{q_i q_j e^2}{m_i} \right)^2 \left(\frac{m_i}{k T_i}\right)^{3/2} \ln\Lambda_{ij},
\end{equation}
with  $\Lambda_{ij}$ as the Coulomb logarithm for ion--ion collisions
\begin{equation}\label{g26-31}
\ln \Lambda_{ij} = \ln  \Lambda_{ji} = 23 - \ln \left[ \frac{q_i q_j (m_i + m_j)}{m_i T_j + m_j T_i} \left(\frac{n_i q_i^2}{T_i} + \frac{n_j q_j^2}{T_j} \right)^{1/2} \right]. \end{equation}

The energy transfer by ion--ion collisions is applied for evaporative cooling where low charged ions cool down more highly charged ions. The the rest gas in the source acts as the source for the LCIs. Because ion--ion collisions occur very efficiently, the ions tend to thermal equilibrium. The energy that the ions have on leaving the ion trap is then proportional to the ion charge state. It is important to mention that there exists an upper limit for the neutral gas density because otherwise the number of highly charged ions will be reduced dramatically by charge exchange processes.

The energy that is transferred out of the ion source by the evaporation process can be described by the cooling rate
\begin{equation}\label{g26-32}
 \displaystyle\frac{\rmd E_0}{\rmd t} \left( \frac{\mbox{eV}}{\mbox{s{\u}cm}^3} \right)  = \displaystyle\frac{\rmd n_{\rm LCI} \, (\mbox{cm}^{-3})}{\rmd t} \, U_{\rm b} \, (\mbox{V}) \,  q_{\rm LCI},
\end{equation}
with $n_{\rm LCI}$ the equilibrium density of the LCIs and $q_{\rm LCI}$ the ion charge state of the LCIs.

\subsubsection{Electron beam: equation of motion and beam radius}

The equation of motion for the electron beam in the paraxial approximation can be written as
\begin{equation}\label{eb-01}
\frac{\rmd^2r}{\rmd t^2} = \frac{e I_\rme}{2\pi\varepsilon_0 v_z r m_\rme} + \frac{e^2}{4m_\rme^2} \left( \frac{B_{\rm c}^2 r_{\rm c}^4}{r^3} - B_z^2 r\right),
\end{equation}
with $B_{\rm c}$ the $B$ field at the cathode, $B_z$ the axial magnetic field, $r_{\rm c}$ the cathode radius, $I_\rme$ the electron beam current  and $v_z$ the axial electron velocity.

If we assume $B_{\rm c}=0$ then there exists a stationary solution of Eq.~(\ref{eb-01}). The solution corresponds to an equilibrium flow of electrons with constant radius, the so-called Brillouin flow (index `B'). For a Brillouin flow, all electrons have a constant distance to the beam centre. Thereby the Lorentz force caused by the magnetic field is compensated by the space charge and the centrifugal force of the rotating electrons.

If we consider more realistic assumptions, such as
\begin{itemize}
\item a magnetic field at the cathode,
\item thermal effects at the cathode due to filament heating up to the temperature $T_{\rm c}$ and
\item interactions between the electrons,
\end{itemize}
then this leads to a corrected electron beam radius (smaller than $r_{\rm B}$) as described by Herrmann \cite{bib:Herrmann1958}.
According to Herrmann's theory the electron beam radius enclosing 80\% of the beam is calculated \cite{bib:Kuramoto2002} to be
\begin{eqnarray}
r_{\mathrm{e}} & = & r(0) \sqrt{\left(1-\frac{r_{0}}{r(0)}\right)^{2}+\frac{2}
{1+({B_{\mathrm{c}}^{2}r_{\mathrm{c}}^{4}}/{B_{{z}}^{2}r_{0}^{4}})}
\left(\frac{v_{\mathrm{e}}\tan\gamma}
{{(e/m_{\mathrm{e}})}B_{{z}}\gamma}\right)},
\label{eq:Herrmannradius}
\end{eqnarray}
and
\begin{eqnarray}
r_{0} & = & r_{\mathrm{B}}\left\{\frac{1}{2}+\frac{1}{2}
\left[1+4\left(\frac{8kT_{{\rm c}}r_{\mathrm{c}}^{2}m_{\mathrm{e}}} {e^{2}B_{{z}}^{2}r_{\mathrm{B}}^{4}}
+\frac{B_{\mathrm{c}}^{2}r_{\mathrm{c}}^{4}} {B_{{z}}^{2}r_{\mathrm{B}}^{4}}\right)\right]^{1/2}\right\}^{1/2},
\label{eq:Herrmann-r_0}
\end{eqnarray}
with $r(0)$ the beam radius at the cathode and  $\gamma$ the angle deviation from the source axis.

\begin{figure}[h]
\centering\includegraphics[width=0.75\linewidth]{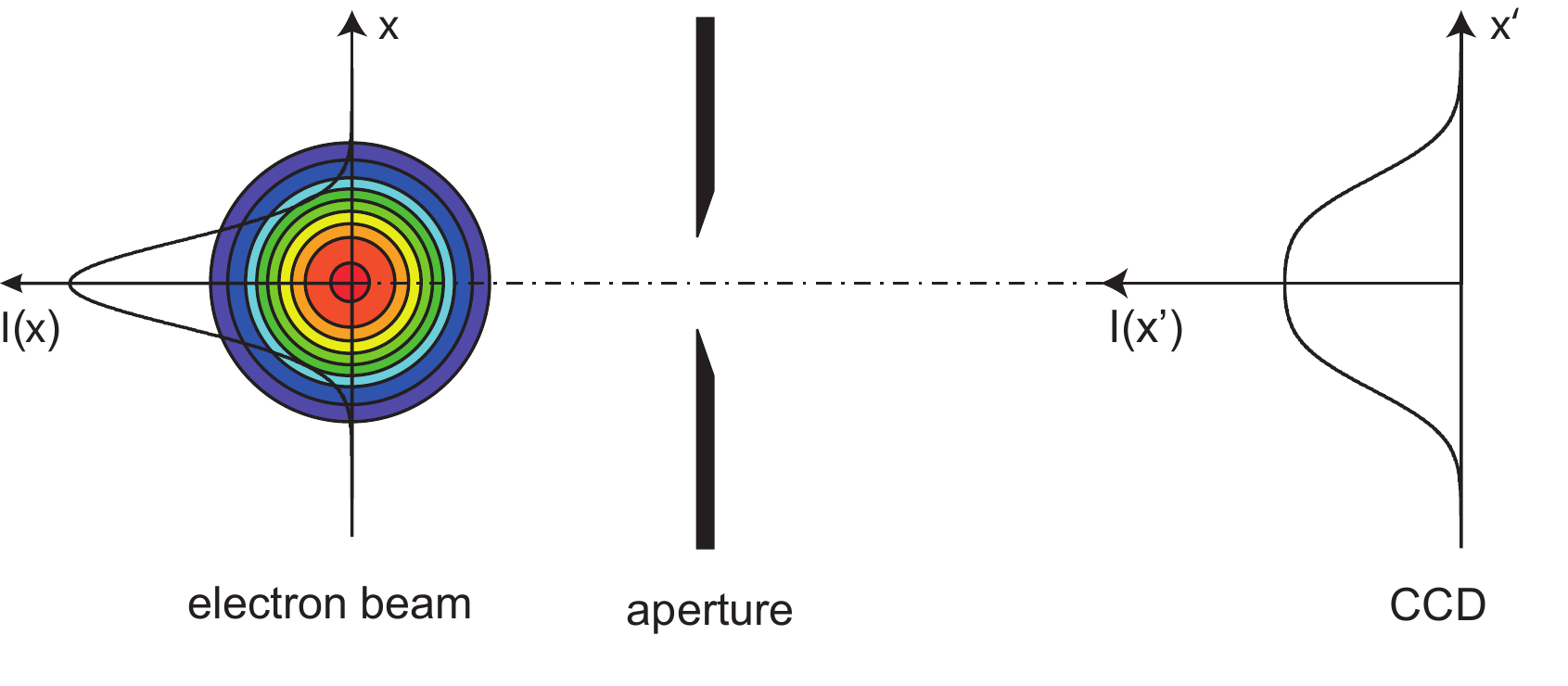}
\caption{CCD chip imaging of X-rays emitted from the electron beam  (after Ref.~\cite{bib:Thorn2012}). Here $I(x)$ is the X-ray intensity distribution from the electron beam, and $I(x^\prime )$ is the intensity distribution on the CCD.}
\label{fig:debeam}
\end{figure}

Generally, the electron beam diameter of an EBIT depends on the actual technical solution. Typical diameters are of the order of 40{\u}$\rmmu$m up to about 200{\u}$\rmmu$m. A classical method to measure the electron beam diameter in an EBIS is shown in Fig.~\ref{fig:debeam} and is discussed in more detail in Ref.~\cite{bib:Thorn2012}. The idea is to measure the X-rays emitted from ions in the electron beam that are ionized or excited by electron impact. Through an aperture the X-rays are detected on a charge-coupled device (CCD) chip and the evaluation of the measured intensity distributions allows the electron beam diameter to be determined.

\section{Ion extraction from EBISs}

\subsection{Basic ion extraction techniques}

As mentioned before, EBISs allow for production of ion beams in DC as well as pulsed mode. In the following, we will discuss and characterize some basic features of EBIS ion extraction. If ions from an EBIS are applied in different experiments, a basic problem is the separation of an individual ion charge state from the extracted ions, which primarily contain a spectrum of individual ion charge states. Generally, the ion charge-state separation is realized by
\begin{itemize}
\item magnetic dipole analyzer or
\item Wien filters.
\end{itemize}

\subsection{Ion beam separation using an analysing magnet}

\begin{figure}[htbp!]
\centering
\includegraphics[width=0.8\linewidth]{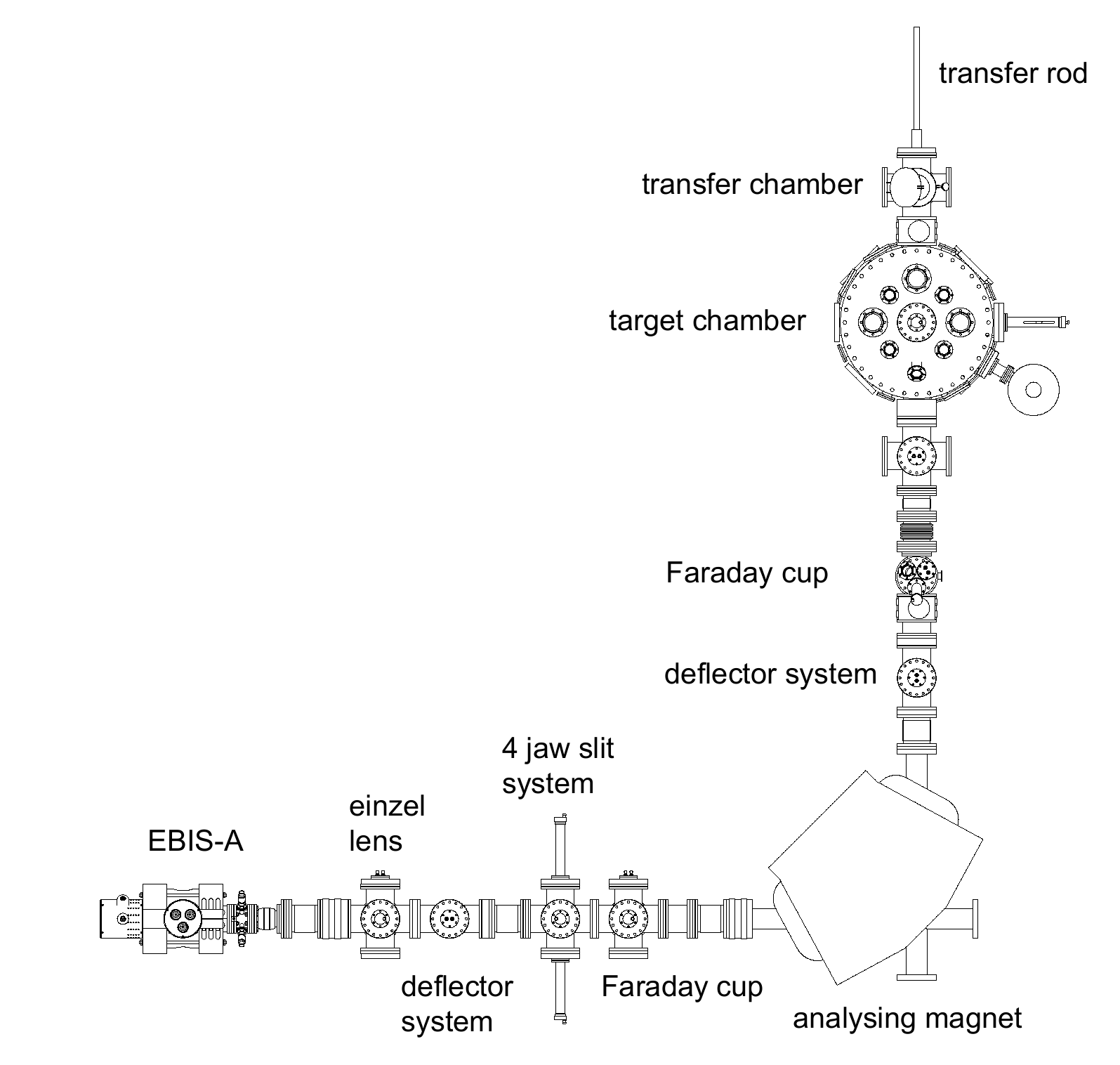}
\caption{Footprint of an ion beam facility featuring a Dresden EBIS-A including a 90$^\circ$ bending magnet for mass-to-charge ratio separation of the ion beam extracted from the EBIS.
\label{fig:MBF}}
\end{figure}

A typical experimental set-up for experiments with extracted ions and magnetic analysis of the ion charge states is shown in Fig.~\ref{fig:MBF}.
The facility shown was designed for experiments with highly charged ions for ion--solid interactions, atomic physics, biophysics, plasma physics, materials analysis, solid-state physics, etc. The ion source as well as the ion beamline can be insulated from ground potential. Thus ions with a wide range of kinetic energies (decelerated as well as accelerated) can be used for experiments in the target chamber of the facility, which can be subdivided into four functional units, the ion source (e.g.\ Dresden EBIS-A), the first and second beamline segments, and the target chamber.

An einzel lens and an $x$--$y$ deflector as well as an aperture, a four-jaw slit system and a Faraday cup are integrated in the first beamline segment. The einzel lens and the $x$--$y$ deflector are used to focus, deflect and align the ion beam.
An aperture with variable diameters can be used, reducing the phase space of the ion beam and therefore improving the beam quality at the expense of beam intensity. A Faraday cup is used for detecting and measuring the ion beam current. It is mounted onto a translational feedthrough directly behind the four-jaw slit system, which is positioned in the focal point of the dipole magnet. Thus the determination of the beam position as well as the beam size is possible.

The ion beam extracted from the ion source can consist of ions of various elements and ion charge states. Thus a dipole magnet separates the ions according to their momentum in the second beamline segment. The momentum of the ion depends on the charge state and the mass of the ion. Hence ions with a certain charge-to-mass ratio  can be selected.

The momentum-separated ions are aligned and deflected with a second $x$--$y$ deflector positioned behind the dipole magnet. A second Faraday cup mounted in the focal point downstream of the dipole magnet ensures detection and current measurement of the ion beam using an electrometer.

The last functional unit of the facility consists of a deceleration lens and the target chamber. The ions are focused into the target chamber by the deceleration lens system. In principle, it is possible to decelerate the ions down to a few electronvolts as well as to accelerate them up to some tens of keV (40{\u}keV) per charge state.
In the example shown, the target chamber has two stages, an upper one for target irradiation experiments and a lower one for target preparation and target transfer. The upper stage is mounted onto the deceleration lens system. The target can be rotated (360$^\circ$) and transferred via a sample holder mounted to a four-axes manipulator. The target can be heated by resistance heating as well as electron bombardment and cooled down via liquid nitrogen. With a `wobble stick' a further target manipulation inside the target chamber is possible. For measuring the beam intensity in the target chamber, a third Faraday cup is mounted onto the four-axes manipulator below the sample holder. A large number of vacuum ports allow the installation of various devices for experiments.

\begin{figure}[htbp!]
\centering
\includegraphics[width=0.60\linewidth]{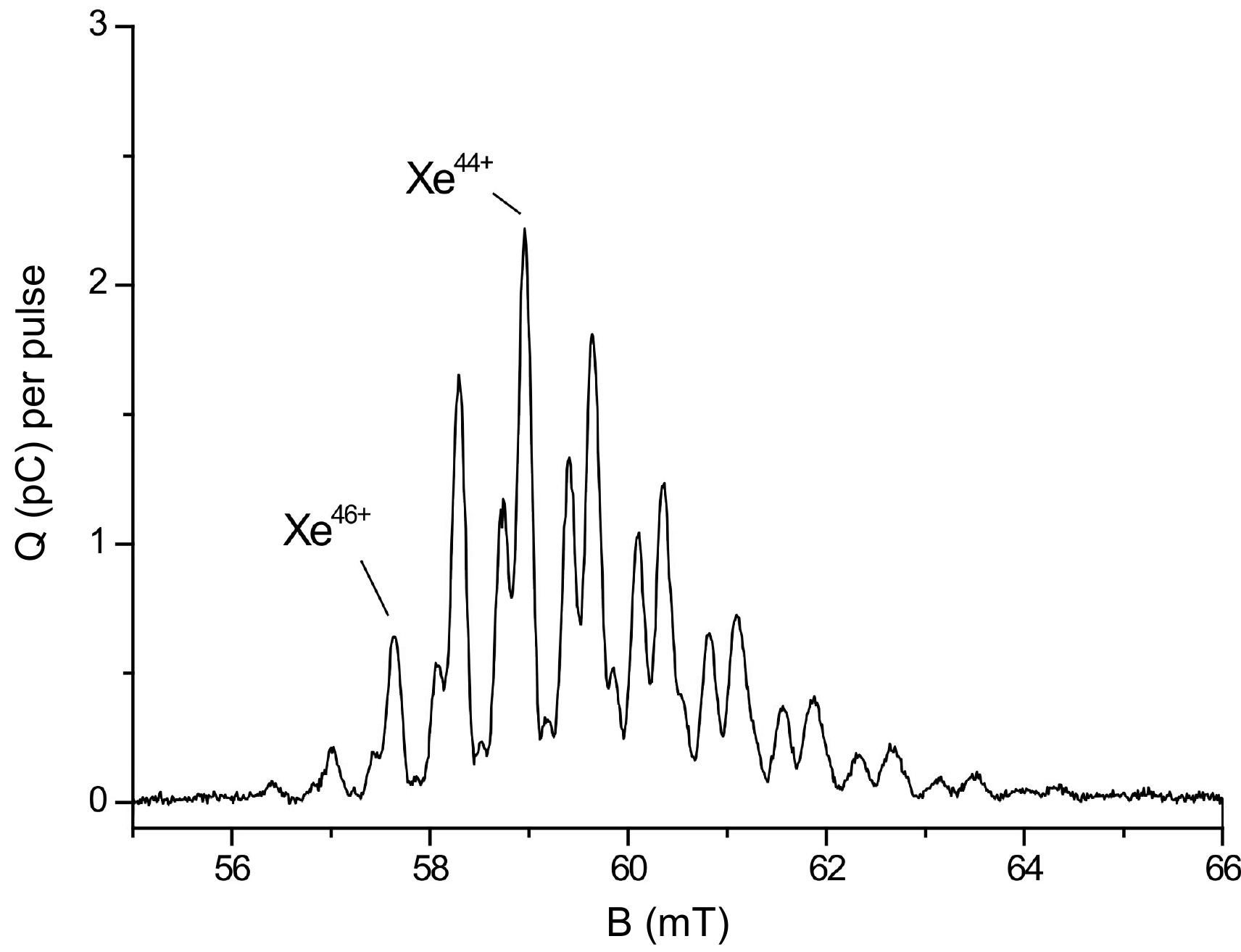}
\caption{A xenon ion extraction spectrum measured at a Dresden EBIS-A for a vacuum of $5\times10^{-10}${\u}mbar, an electron beam energy of 12{\u}keV and an electron beam current of 100{\u}mA.   \label{fig:xe44+}}
\end{figure}

\begin{figure}[htbp!]
\centering
\includegraphics[width=0.60\linewidth]{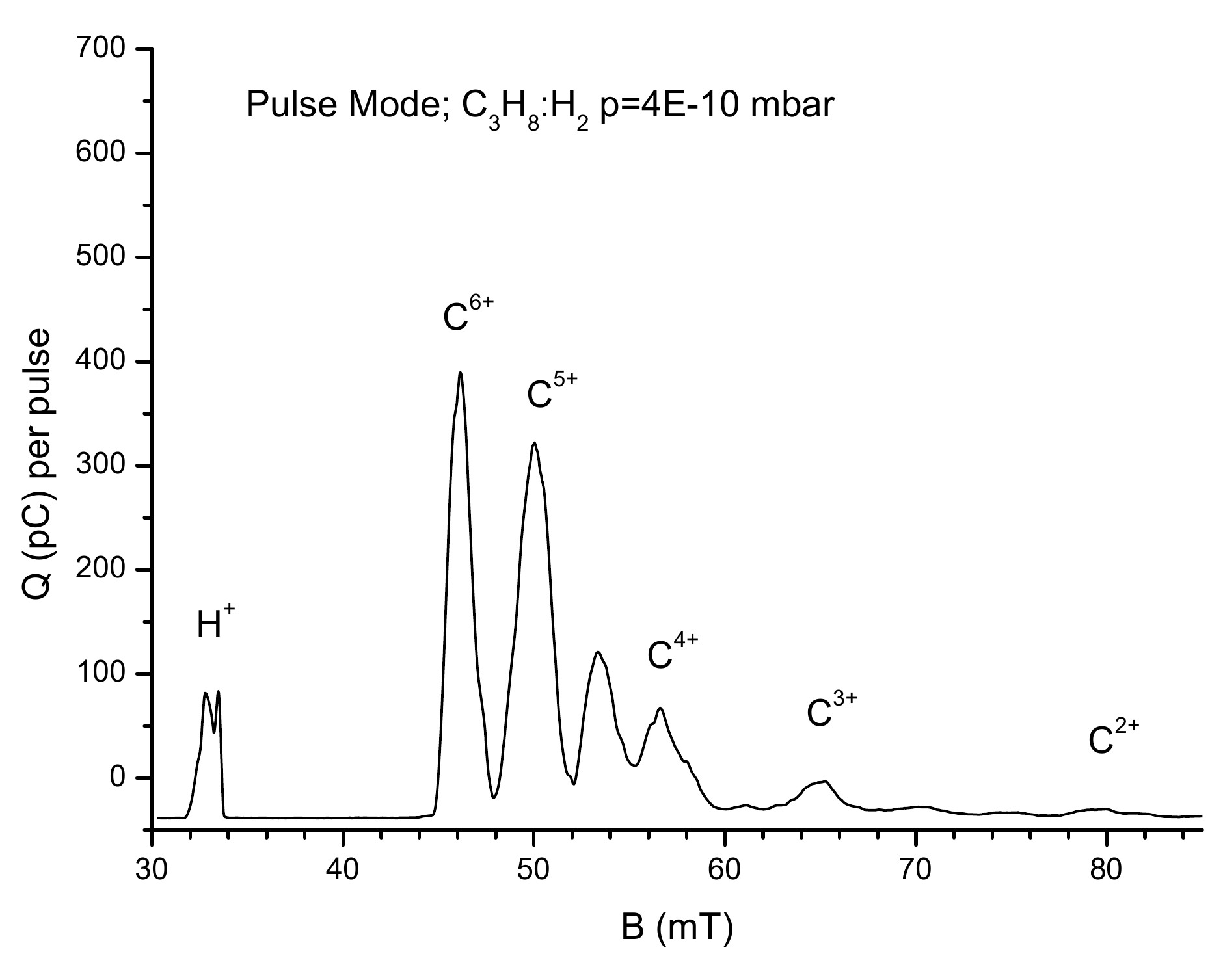}
\caption{A carbon ion extraction spectrum measured at a Dresden EBIS-SC for a vacuum of $4\times10^{-10}${\u}mbar, an electron beam energy of 12{\u}keV, an electron beam current of 400{\u}mA and an ionization time of 200{\u}ms.   \label{fig:c6+-100ms}}
\end{figure}

To characterize ion extraction spectra with an analysing magnet, some examples for ion extraction spectra are given as they can be expected for EBISs. Figure \ref{fig:xe44+} shows a xenon ion extraction spectrum as measured with a Dresden EBIS-A for an ionization time of 5{\u}s. In Fig.~\ref{fig:c6+-100ms} a carbon ion extraction spectrum is shown as measured with a superconductiong Dresden EBIS-SC for an ionization time of 200{\u}ms (for details see Ref.~\cite{Zschornack2012}).

\subsection{Ion beam separation using a Wien filter}

\begin{figure}[htbp!]
\centering
\includegraphics[width=0.60\linewidth]{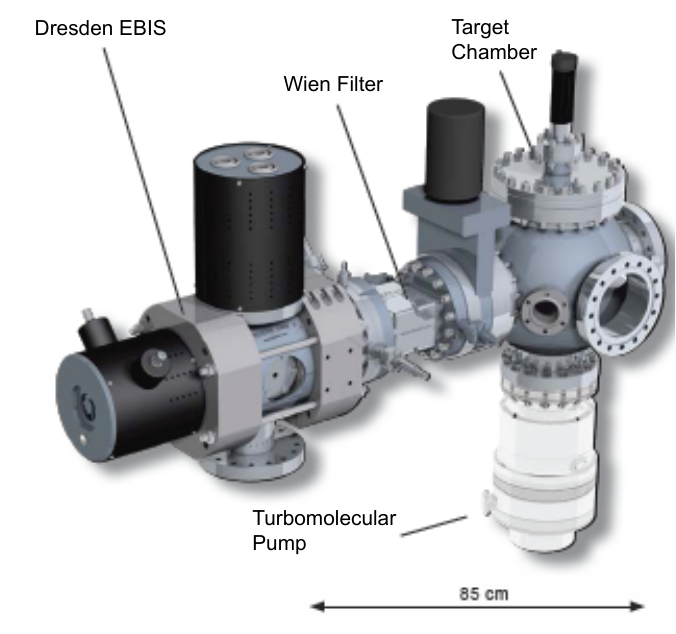}
\caption{A 3D presentation of a Wien filter-based beamline
\label{fig:fac-s}}
\end{figure}

A Wien filter is a particle separator with crossed
magnetic and electric field configuration providing
mass- and charge-state-separated beams of excellent
quality and small spatial dimensions  \cite{bib:Schmidt2009}.
With the Wien filter as an ion source add-on, a very
compact device is available substituting a complete
standard beamline set-up with a space-consuming
dipole magnet and other necessary equipment (see Fig.~\ref{fig:fac-s}).
The Wien filter can be used as a stand-alone solution
for various beamline set-ups. Depending on the
installed mass- and charge-state-separating aperture,
a resolution of better than 80 is available.
The geometrical dimensions of the Wien filter are small
compared to a classical dipole magnet. The Wien filter
fits easily on a beamline, making a change of the ion
beam direction, as is necessary for a dipole
magnet, obsolete.

\begin{figure}[htbp!]
\centering
\includegraphics[width=0.60\linewidth]{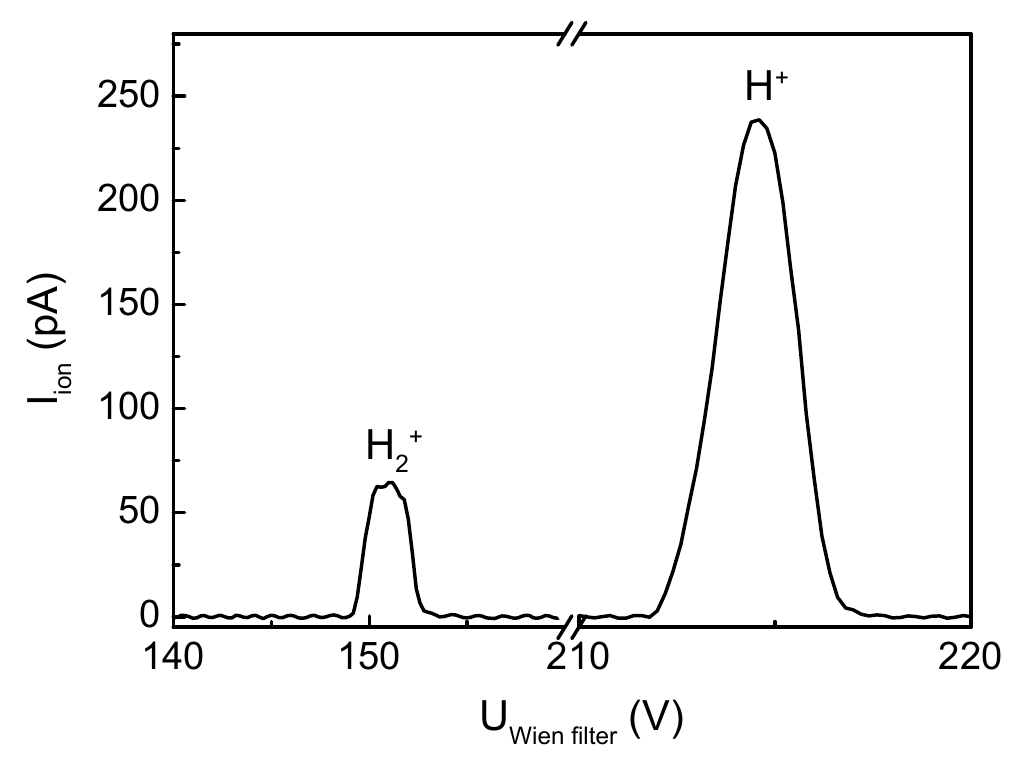}
\caption{A leaky mode hydrogen ion extraction spectrum measured at a Dresden EBIS for a vacuum of $3\times10^{-9}${\u}mbar, an electron beam energy of 13.6{\u}keV and an electron beam current of 30{\u}mA.
\label{fig:wf-h}}
\end{figure}

\begin{figure}[htbp!]
\centering
\includegraphics[width=0.70\linewidth]{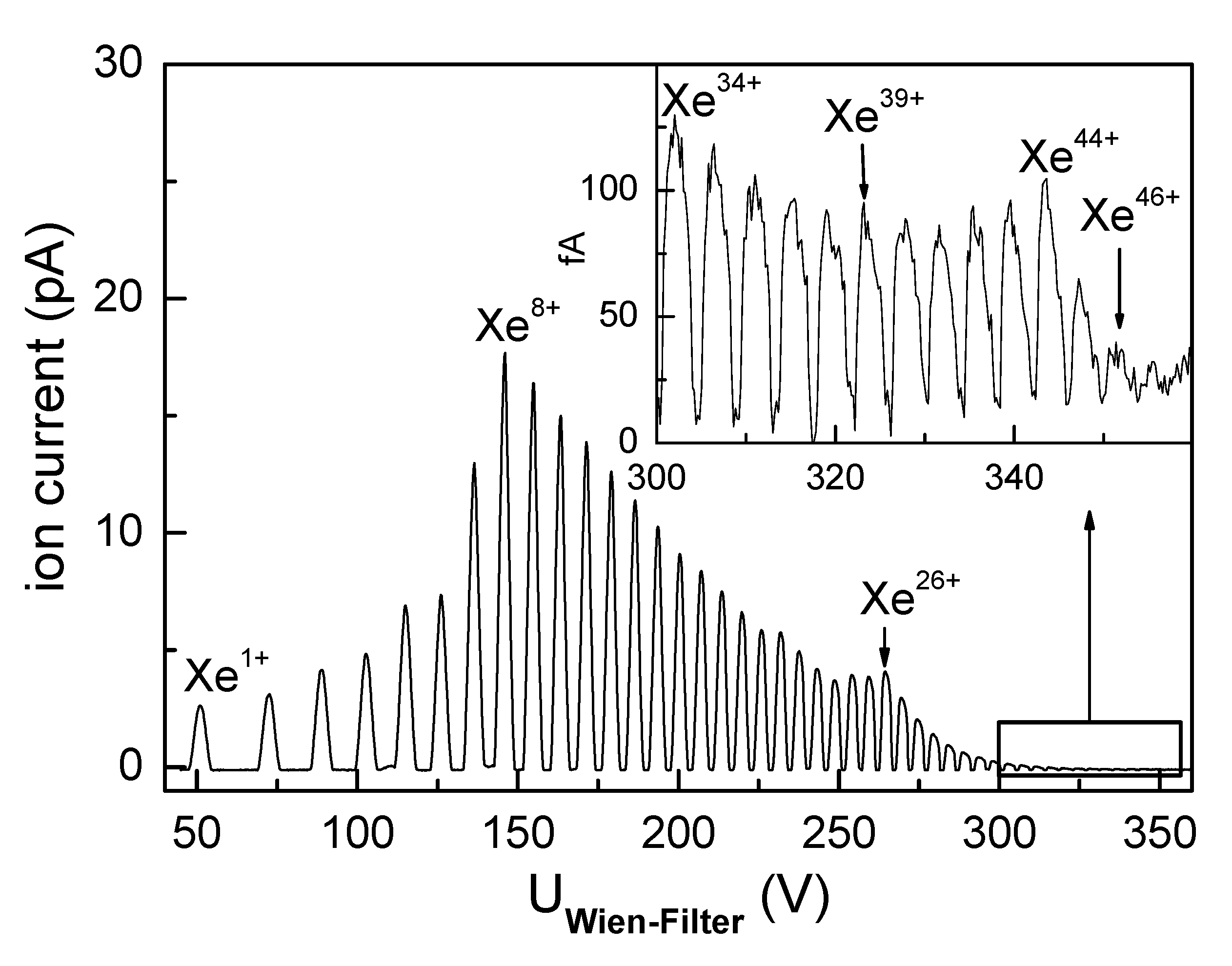}
\caption{A leaky mode xenon ion extraction spectrum measured at a Dresden EBIS for a vacuum of $3\times 10^{-9}$~mbar, an electron beam energy of 13.6 keV and an electron beam current of 30 mA.   \label{fig:wf-xe}}
\end{figure}

The use of a Wien filter allows for separation of light and intermediate-$Z$ ions of all ion charge states. For hydrogen and xenon, this is shown in Figs.~\ref{fig:wf-h} and \ref{fig:wf-xe}. The most important difference to the use of analysing magnets is that a magnet has a gap of several centimetres, but for obtaining high resolution with a Wien filter we must work with aperture diameters in the millimetre or sub-millimetre region, i.e.\ beam losses through the Wien filter have to be taken into account.

\subsection{Beams of singly charged molecular fragments and clusters}

\begin{figure}[htbp!]
\centering
\includegraphics[width=0.85\linewidth]{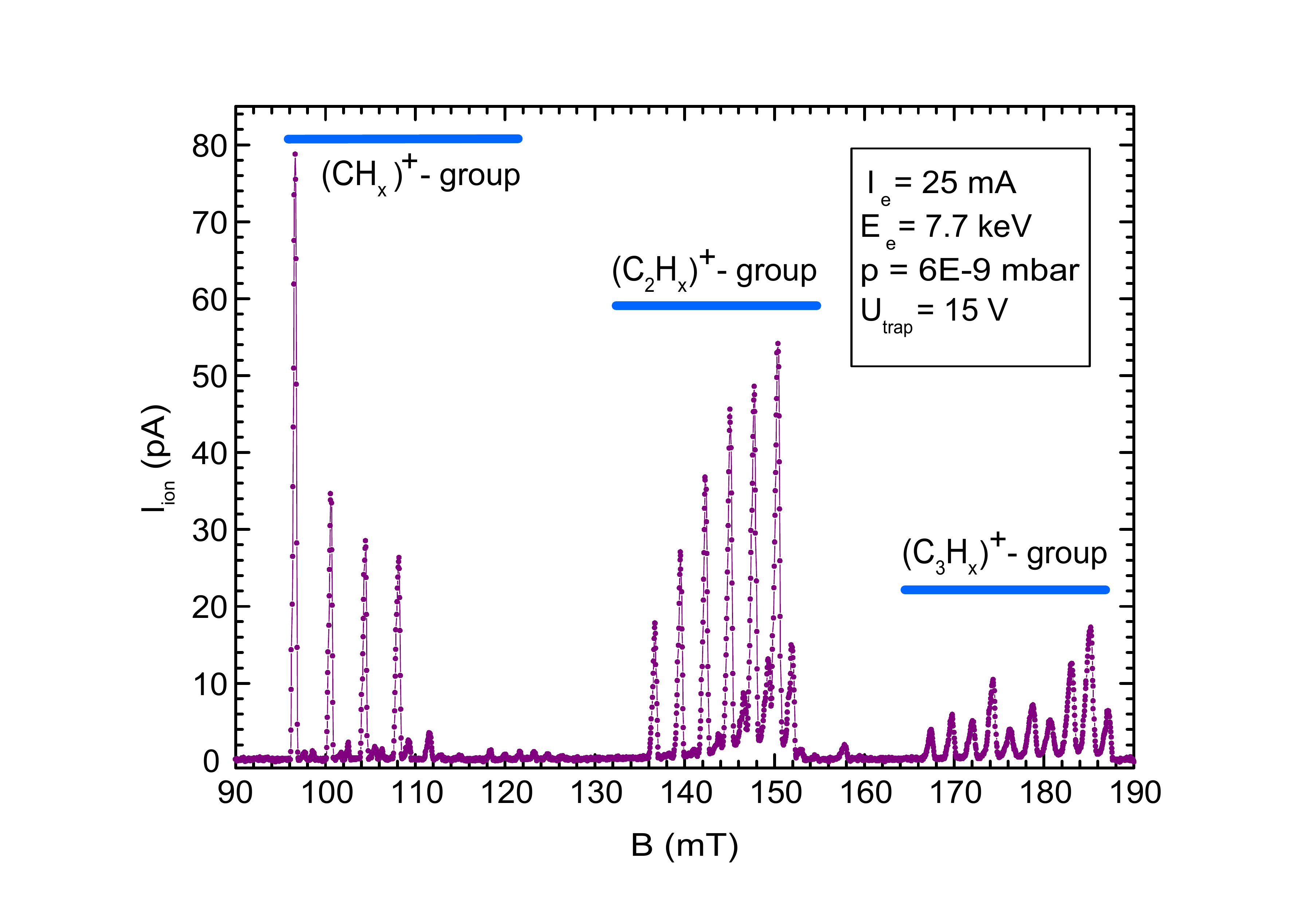}
\caption{Molecular fragment spectrum in leaky mode extracted from the Dresden EBIT after injection of propane  \label{fig:propane}}
\end{figure}

In leaky mode, EBISs can be used as sources for low charged ions from molecular fragments. The production of molecular fragments in an EBIS is of interest for various applications. This concerns issues such as studies on electronic transitions in molecules by spectroscopy of emitted electromagnetic radiation, protonation studies on complex molecules, as well as investigations on the interaction of molecular fragments with surfaces and on free molecules or gases in a corresponding gas jet. The use of a gas target may allow the production of neutral beams of different molecular fragments by charge exchange processes. Figure \ref{fig:propane} shows an ion extraction from a Dresden EBIT after injection of propane. Thereby all possible stoichiometric ratios of propane fragments were detected. More details about molecule fragmentation in EBISs can be found in Ref.~\cite{bib:Kreller2008}.

\begin{figure}[htbp!]
\centering
\includegraphics[width=0.50\linewidth]{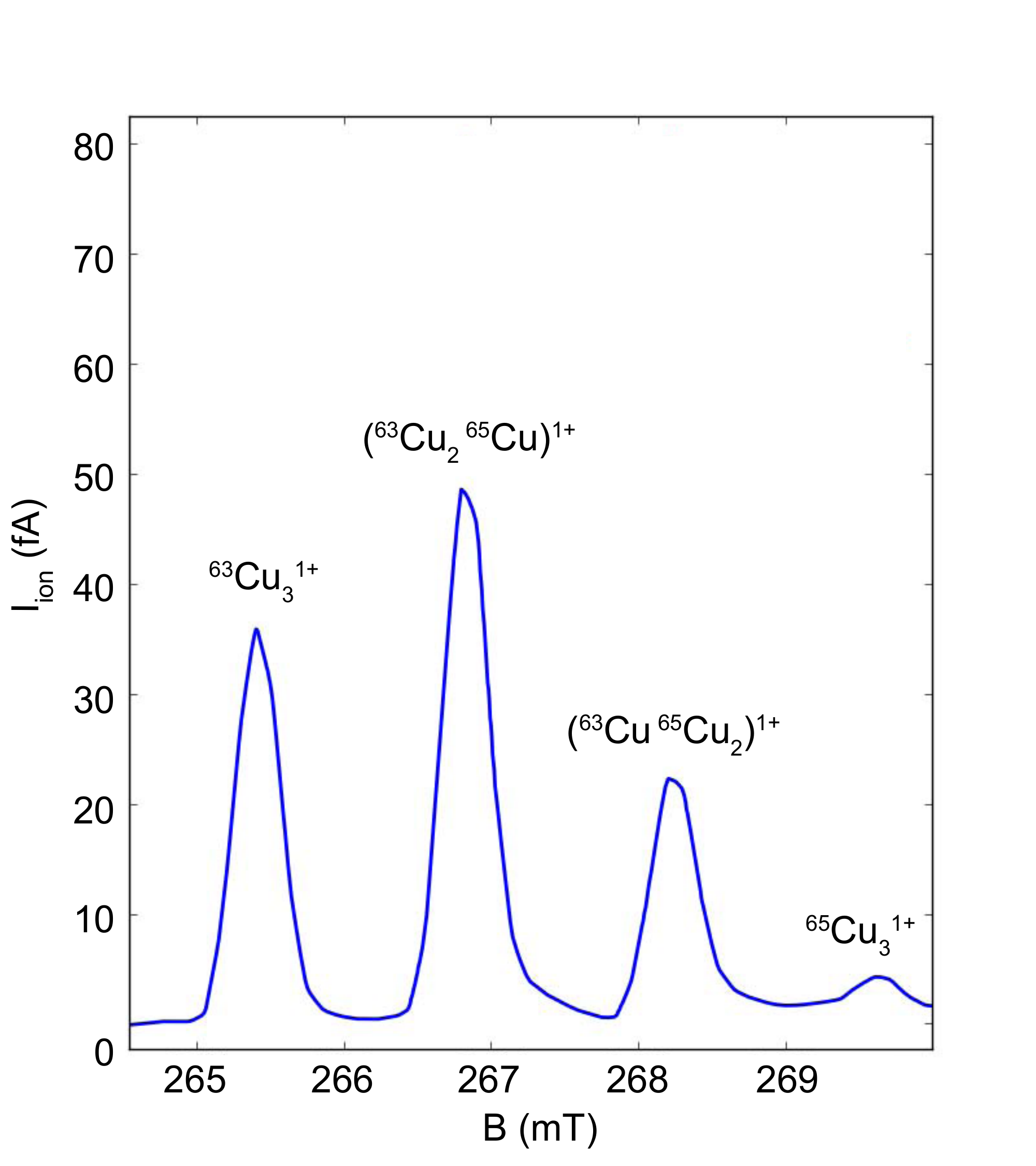}
\caption{Measured copper clusters from a Dresden EBIT at an electron beam energy of 6{\u}keV and an electron beam current of 23{\u}mA.  \label{fig:cu-cluster}}
\end{figure}

Additionally, one can also produce more complex structures such as metal clusters. In Fig.~\ref{fig:cu-cluster} this is shown for copper clusters as produced with a Dresden EBIT. The copper was introduced into the ion source by sputtering from a copper surface under gold ion irradiation. The  capability of this technique is seen if we note that clusters with different isotope ratios can be clearly resolved, such as
[($^{63}$Cu)$_3$]$^{1+}$, [($^{63}$Cu)$_2$($^{65}$Cu)$_1$]$^{1+}$, [($^{63}$Cu)$_1$($^{65}$Cu)$_2$]$^{1+}$ and [($^{65}$Cu)$_3$]$^{1+}$. Without a great deal of effort, cluster beam currents of up to $10^5$ clusters per second were measured.

\subsection{Absolute numbers of extracted ions}

To characterize the potential of different EBIS types, we give some absolute ion numbers for pulsed ion extraction of ion sources of the Dresden EBIS type in Table \ref{tab:ionnumbers}. The summarized numbers characterize the expected order of magnitude for ion extraction but can deviate in their dependence on the individual ion source and the selected operation regime.

\begin{table}[htb]
\caption{Ions per pulse for pulsed ion extraction of ion sources of the Dresden EBIS type. \label{tab:ionnumbers}}
\centering
\begin{tabular}{lcccc}
 \hline\hline
 \noalign{\vspace{4pt}}
 Ion & Dresden EBIT & Dresden EBIS & Dresden EBIS-A & EBIT : EBIS : EBIS-A\\[1ex]
 \hline
 \noalign{\vspace{6pt}}
 C$^{6+}$ & 10\,000\,000 & 30\,000\,000 & 400\,000\,000 & 1 : 3 : 40\\[1ex]
 Ar$^{16+}$ & 900\,000 & 8000\,000 & 250\,000\,000 & 1 : 9 : 280 \\[1ex]
 Ar$^{18+}$ & 6000 & 90\,000 & 1500\,000 & 1 : 15 : 250\\[1ex]
 Xe$^{44+}$ & 10\,000 & 700\,000 & 10\,000\,000 & 1 : 70 : 1000 \\[1ex]
 \hline\hline
\end{tabular}
\end{table}

\subsection{Pulse shape for typical ion extraction}

\begin{figure}[htbp!]
\includegraphics[width=0.55\linewidth]{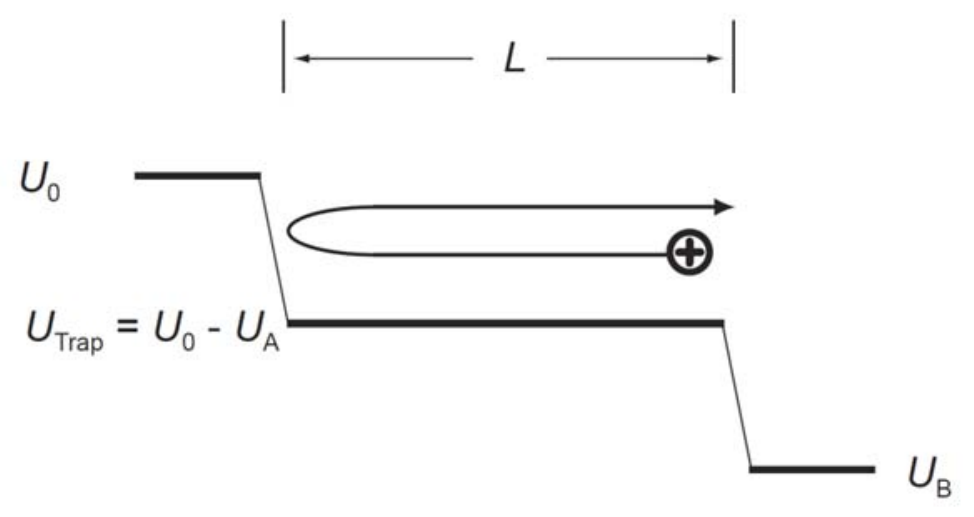}
\centering
\caption{Ion paths at the moment of ion extraction
\label{fig:EBIS-pulsformation}}
\end{figure}

\begin{figure}[htbp!]
\includegraphics[width=0.6\linewidth]{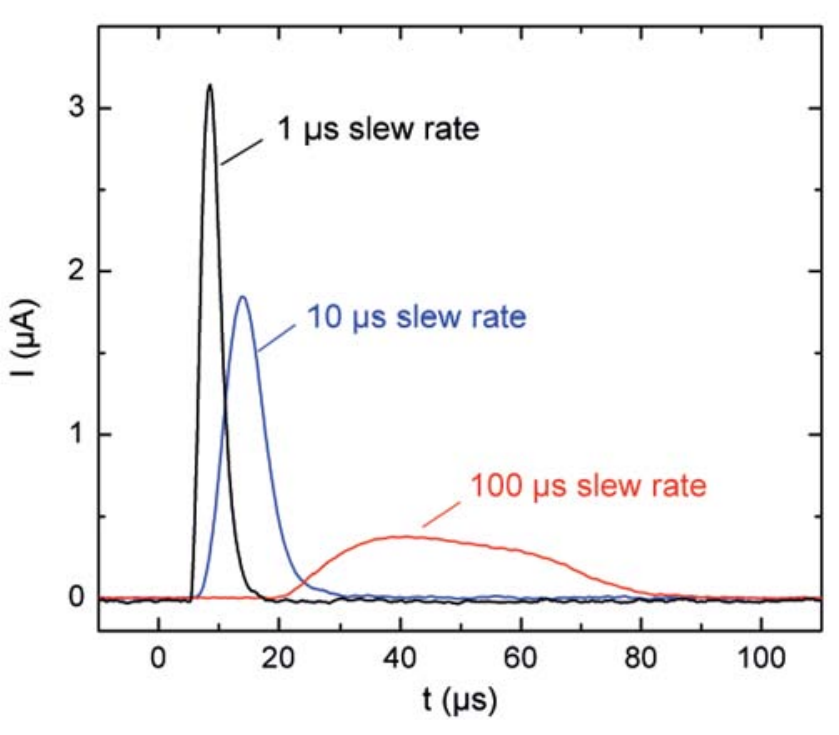}
\centering
\caption{Carbon ion pulses from a Dresden EBIS-A for different slew rates
\label{fig:flat-top}}
\end{figure}

Typical pulse widths at pulsed ion extraction from an EBIS are of the order of a few microseconds. Thereby the pulses are not completely symmetric but have a tail on the side of higher flight times. The reason is that, at the moment of trap opening, only those ions are extracted that have a flight direction towards the extraction side. All other ions will be reflected at the potential wall at the gun side and extracted after that. This situation is shown in Fig.~\ref{fig:EBIS-pulsformation}. The absolute pulse width is determined by the speed of switching the trap potential at the extraction side (`slew rate'). We see that we can influence the pulse width from some microseconds up to about 100{\u}$\rmmu$s, i.e.\ we come near to `flat-top' pulses. As an example this is shown in Fig.~\ref{fig:flat-top} for ion pulses from a Dresden EBIS-A with a trap length $L$ of 6 cm and different slew rates.

\subsection{Short-time ion pulse extraction}

\begin{figure}[htbp!]
\centering
\includegraphics[width=0.6\linewidth]{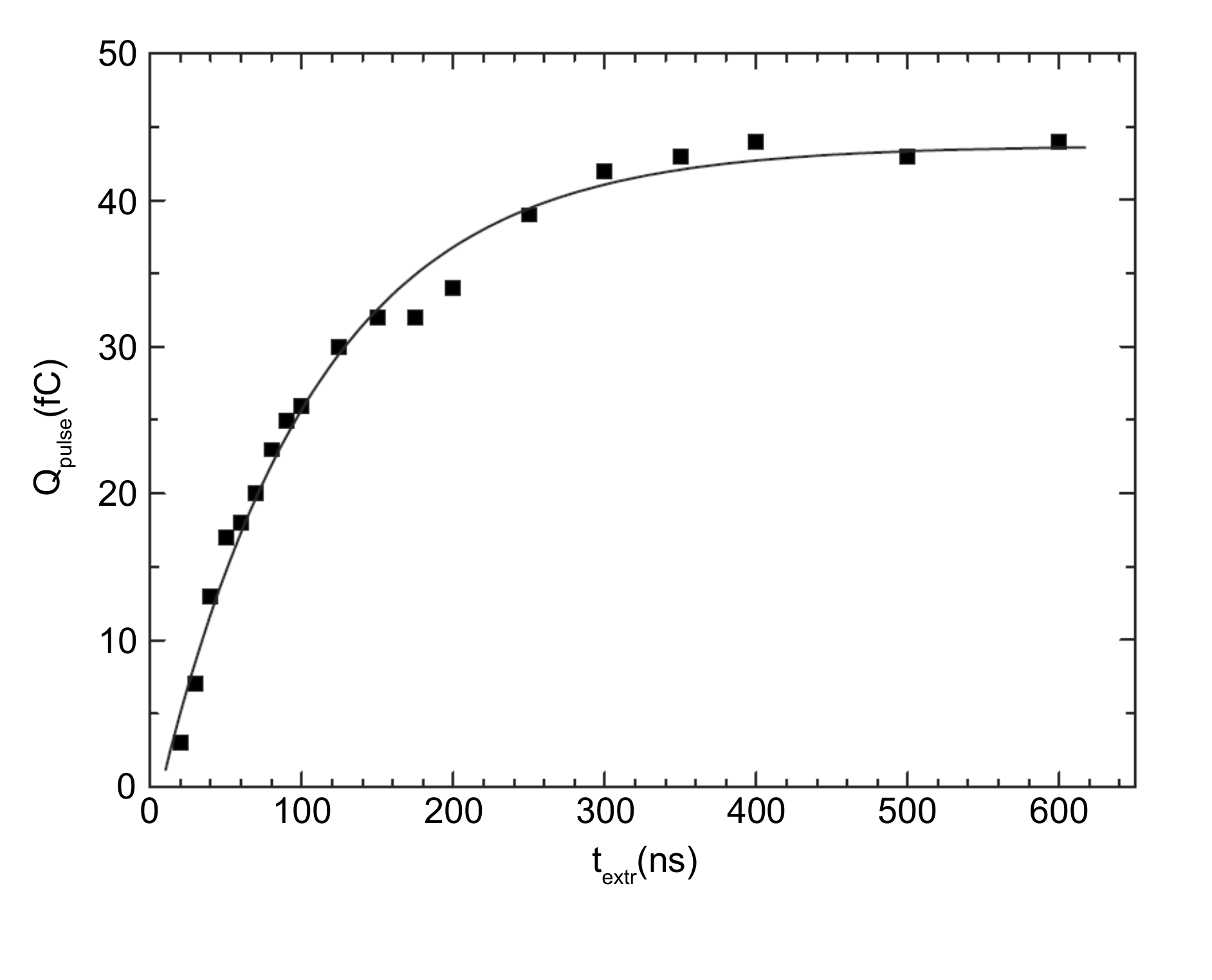}
\caption{Extracted ion charges of Ar$^{16+}$ pulses from a Dresden EBIT as a function of the extraction time $t_{\rm extr}$ (after Ref.~\cite{bib:Kentsch2010}).  \label{fig:ar-ns}}
\end{figure}

For some applications the time structure of the extracted
beams of HCIs is of fundamental interest. For example, for applications  in time-of-flight secondary ion mass spectroscopy (TOF-SIMS) and in some solid-state physics experiments, the formation of short pulses up to the nanosecond range is of interest. Present studies prove that a short pulsed ion extraction
from the Dresden EBIT is possible. The extracted ion
charges per pulse are in the range of some femtocoulombs and it is possible
to extract several pulses until the trap has to be reloaded.
The maximum extracted ion charge depends significantly on the
chosen extraction time. As an example we show in Fig.~\ref{fig:ar-ns} a plot of extracted ion charges for an Ar$^{16+}$ pulse as a function of the extraction time $t_{\rm extr}$. From Fig.~\ref{fig:ar-ns} it is seen that, after some hundreds of nanoseconds, the extracted ionic charge increases no further.

\subsection{Ion beam emittance}

For applications of ion beams, a quantitative measure of the quality of the beam is
given by the root-mean-square (r.m.s.) emittance
\begin{equation}\label{emit-1}
\varepsilon_{x,\rm rms} = \sqrt{\overline{x^2}\,\overline{{x^\prime}^2} - \overline{{x^\prime}^2}}
\end{equation}
in $x\mbox{--}x^\prime$ trace space and equivalently in the case of $y\mbox{--}y^\prime$.

\begin{figure}[htbp!]
\centering
\includegraphics[width=0.7\linewidth]{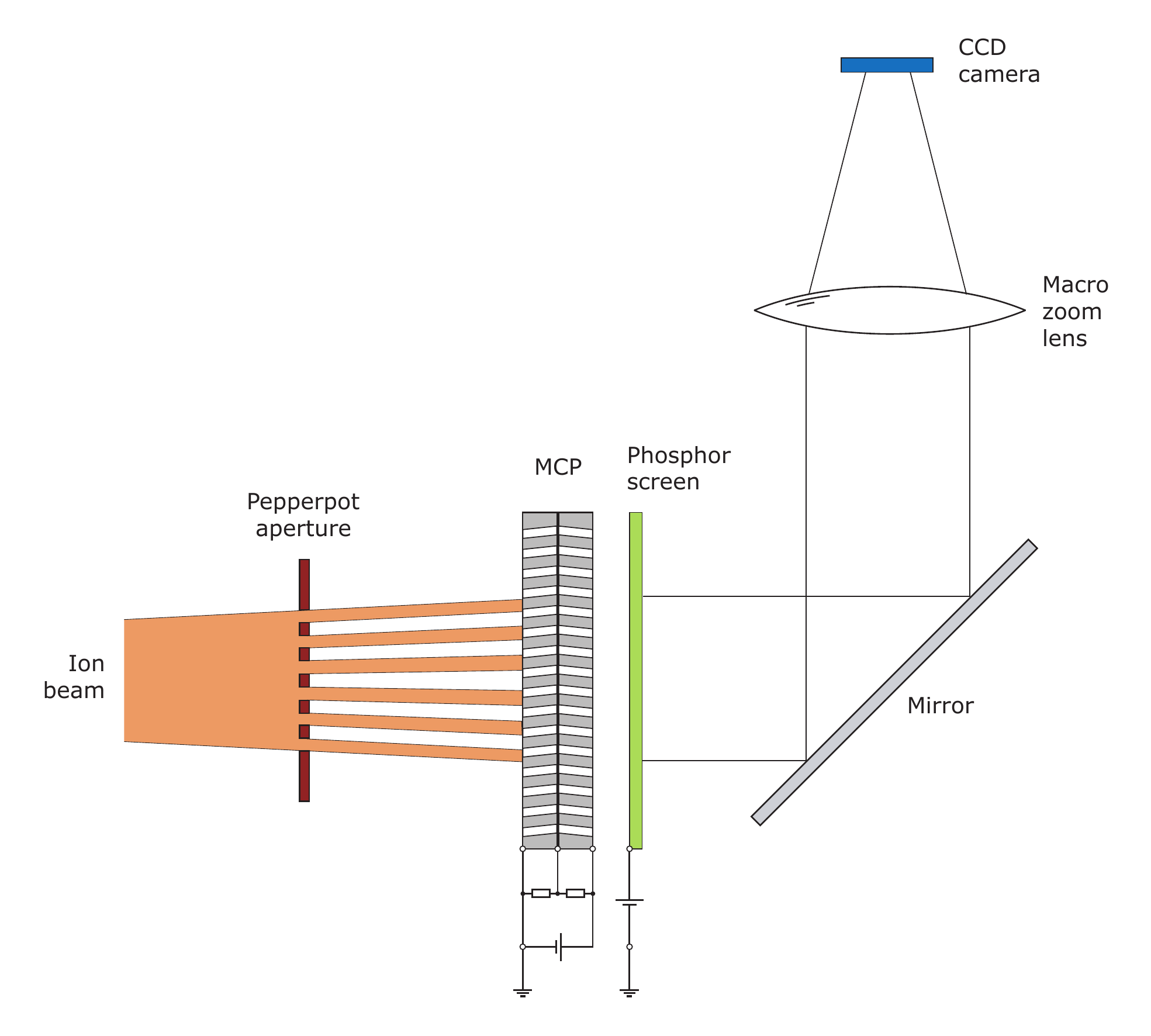}
\caption{Scheme of a pepperpot emittance meter  \label{fig:pepp}}
\end{figure}

A common  method to determine these ion beam parameters is
the pepperpot technique, which can be traced back to works
of the 1960s \cite{bib:Collins1964}. A scheme of a pepperpot emittance meter is shown in Fig.~\ref{fig:pepp}. The incoming ion beam passes the pepperpot mask and is separated into several beam spots. The particles hitting the microchannel plate (MCP) create an electron current, which is amplified passing the two MCPs. The electrons are then accelerated towards the phosphor screen. The visible light spots created at the phosphor screen are detected after 90$^\circ$ deflection by a CCD camera. The emittance of the ion beam can be determined from the positions, sizes and shapes of the light spots.

\begin{figure}[htbp!]
\centering
\includegraphics[width=0.65\linewidth]{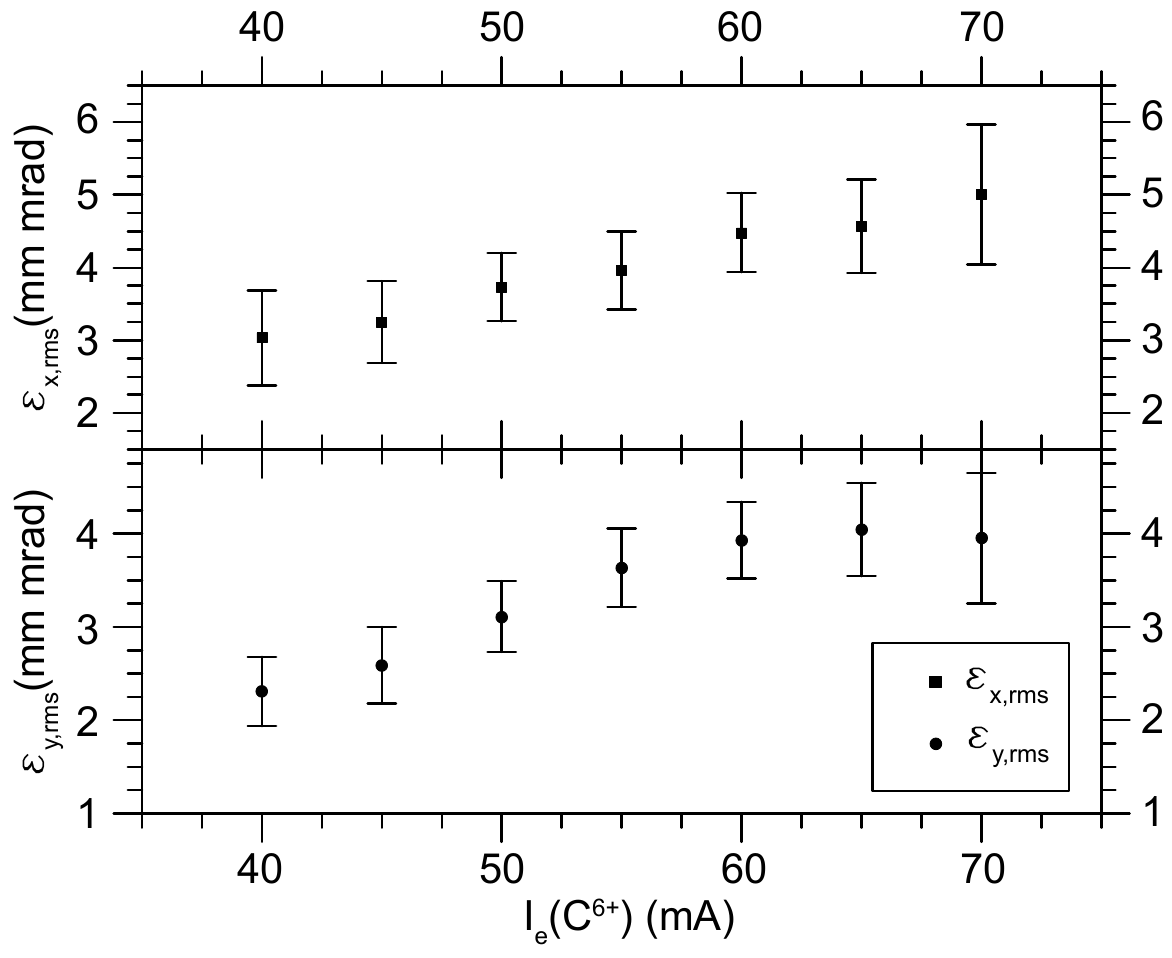}
\caption{R.m.s.\ emittance of C$^{6+}$ ions extracted from a Dresden EBIS-A  \label{fig:emit_c6}}
\end{figure}

Because of the small ionization volume and the small electron beam diameter, EBISs produce ion beams with very low emittances. As an example in Fig.~\ref{fig:emit_c6} the $x$ and $y$ r.m.s.\ emittance for C$^{6+}$ ions extracted from a Dresden EBIS-A is shown as a function of the electron beam current. Typical emittances are of the order of some mm{\u}mrad.

For higher beam energies, the pulse components perpendicular to the beam direction are small compared with the pulse components in the beam direction, i.e.\ higher beam energies lead to lower emittances. This can be described by the normalized emittance and it yields
\begin{equation}\label{emitt-2}
\varepsilon_{\rm norm} = \beta^\prime \gamma^\prime \varepsilon \qquad\quad \mbox{with} \quad \beta^\prime = \frac{v}{c} \quad\mbox{and}\quad 
\gamma^\prime = \frac{1}{\sqrt{1-\beta^2}},
\end{equation}
where $v$ is the ion velocity and $c$ is the speed of light. For typical operation regimes of EBISs, the normalized emittance is of the order of $10^{-3}${\u}mm{\u}mrad.

\subsection{Energy spread and absolute energy of extracted ions}

In order to measure the absolute energy and the energy spread of extracted ions, a retarding field analyser (RFA) can be applied.
An RFA is an electrostatic energy analyser that allows the kinetic energy distribution of an ion beam to be measured. The analyser consists of a collimator with exchangeable apertures, three meshes, with the central mesh set on high potential, and a Faraday cup for current measurement.

The ion or electron current is measured in the Faraday cup of the RFA.
In order to analyse the energy distribution of the incoming ion beam, the potential of the retarding mesh is increased stepwise while measuring the current in the cup. The measured dependence $I_{\rm ion}(U_{\rm RFA})$ is differentiated and fitted with a Gaussian to obtain the energy distribution. The mean energy of the charged particles $E_{\rm m}$ is defined as the centre position of the Gaussian.
The energy spread of the charged particle beam $\Delta E$ is defined as the full width half-maximum (FWHM) of the Gaussian.

\begin{figure}[htbp!]
\centering
\includegraphics[width=0.7\linewidth]{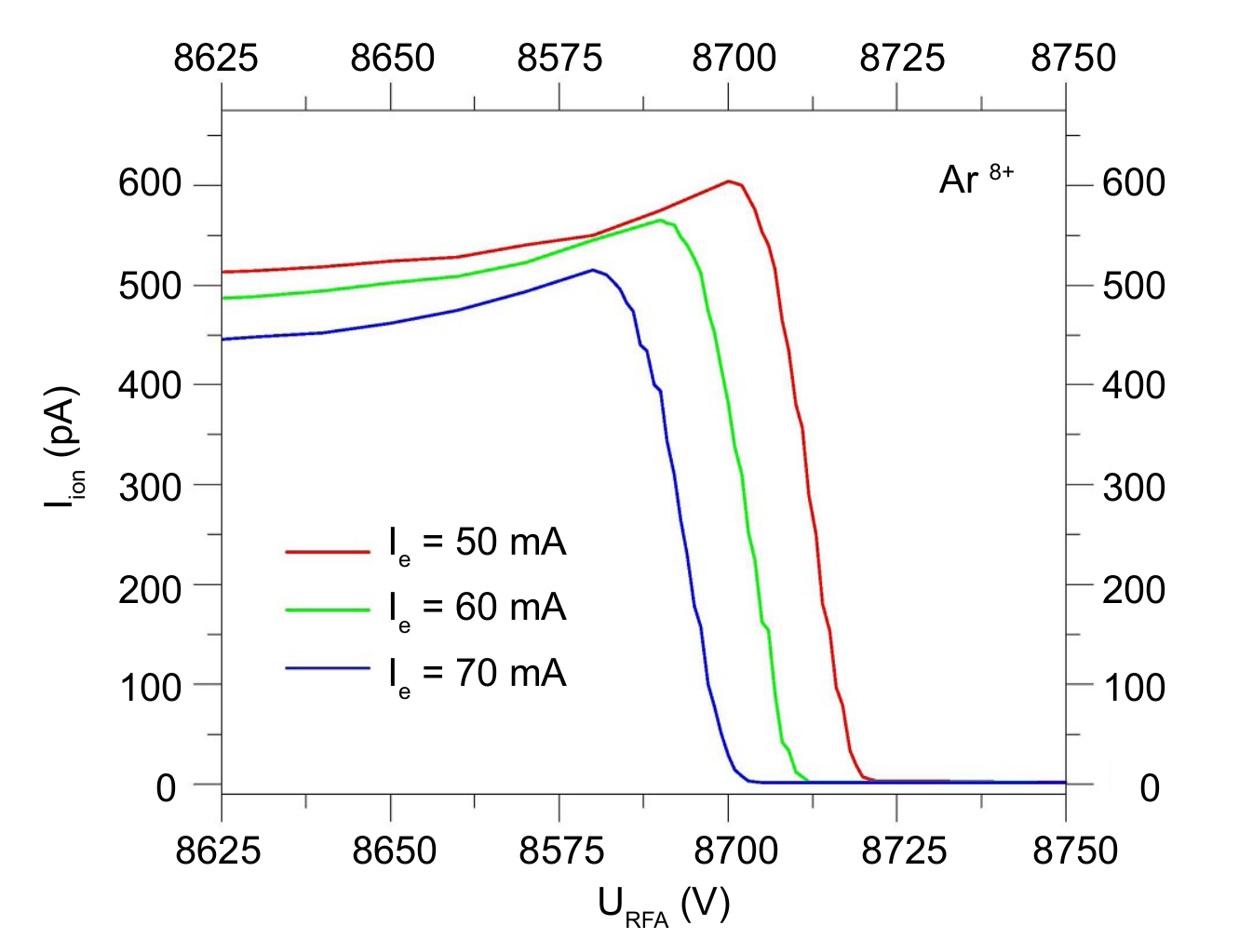}
\caption{Plot showing $I_{\rm ion}(U_{\rm RFA})$ curves for electron currents of 50, 60 and 70{\u}mA measured with Ar$^{8+}$ at a drift tube potential of 8.7{\u}keV  at a Dresden EBIS-A.  \label{fig:RFA_50&60&70mA}}
\end{figure}

\begin{figure}[htbp!]
\centering
\includegraphics[width=0.7\linewidth]{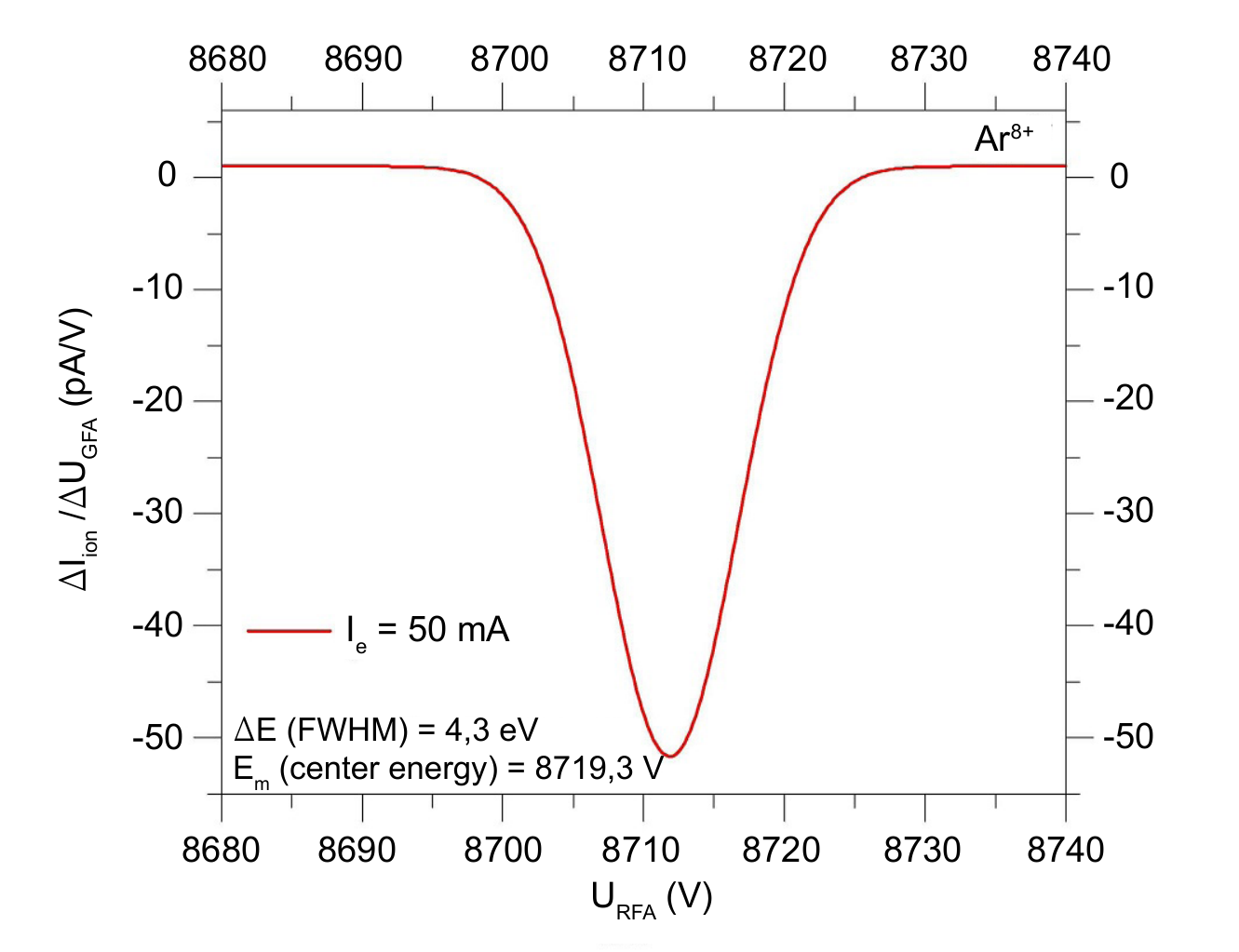}
\caption{The energy distribution of the extracted ion beam is $E = 4.3${\u}eV at an electron beam of $I_\rme = 50${\u}mA measured at a Dresden EBIS-A.  \label{fig:RFA_energy_distribution_50mA}}
\end{figure}

Figures \ref{fig:RFA_50&60&70mA} and \ref{fig:RFA_energy_distribution_50mA} show examples of measured absolute energies and energy spread of ions extracted from a Dresden EBIS-A. The energy spread of the ion beam, which depends on the electron beam current, is generally quite small for Dresden EBIS/T systems. For the Dresden EBIS-A, it is of the order of several electronvolts. More details about the RFA technique can be found in Ref.~\cite{bib:Ritter2010}.

\section{EBISs as sources of X-rays from HCIs}

\subsection{X-ray emission from HCIs}

EBISs are excellent tools for X-ray spectroscopy of highly charged ions. X-rays from HCIs are of fundamental interest for atomic physics, plasma diagnostics, astrophysics and other applications. In contrast to other sources of X-rays from HCIs such as hot plasmas or accelerator facilities, EBISs are comparatively small and low-cost devices. Additionally, they offer the possibility to control the electron beam energy very precisely, which creates superior conditions for advanced experiments studying X-rays from HCIs.

It is well known that vacancies in the atomic shell influence X-ray lines in terms of their transition energies, transition probabilities and the structure of the measured spectrum. Classical X-ray spectra have been known for more than 100 years. As long ago as 1917 Barkla received a Nobel Prize for the discovery of characteristic X-rays. Less than 50 years later, in 1962, the first X-ray source outside our Solar System was observed and from that date interest in X-rays from HCIs has grown more and more.

Compared to classical X-rays as parent diagram lines appearing after the production of an inner-shell vacancy and subsequent filling of this vacancy with an outer electron, X-rays from HCIs can be different from these classical transitions. Mainly, we can observe X-ray transitions from
\begin{itemize}
\item direct excitation (DE),
\item radiative recombination (RR) and
\item dielectronic recombination (DE).
\end{itemize}

Detailed information on the basic physics of X-rays from HCIs can be found, for example, in Refs.~\cite{bib:Beyer1997} and~\cite{bib:Currell2003b}. Measurements were performed by different groups working with EBISs -- the pioneering work of the Livermore group of Dr.\ Beiersdorfer \textit{et al.} should be pointed out especially (see e.g.\ Ref.~\cite{bib:Beiersdorfer1997}).

\subsection{X-ray measurements}

Measurements show that in the case of the Dresden EBIT individual dipole lines have a radiation power of the order of some nanowatts. For the Dresden EBIS-A this power increases by about one order of magnitude and for the superconducting Dresden EBIS-SC by a factor of about 200. As a rule, the number of X-rays emitted is high enough to perform energy-dispersive as well as wavelength-dispersive X-ray spectrometry in acceptable time frames of minutes up to some hours.

\begin{figure}[htbp!]
\centering
\includegraphics[width=0.8\linewidth]{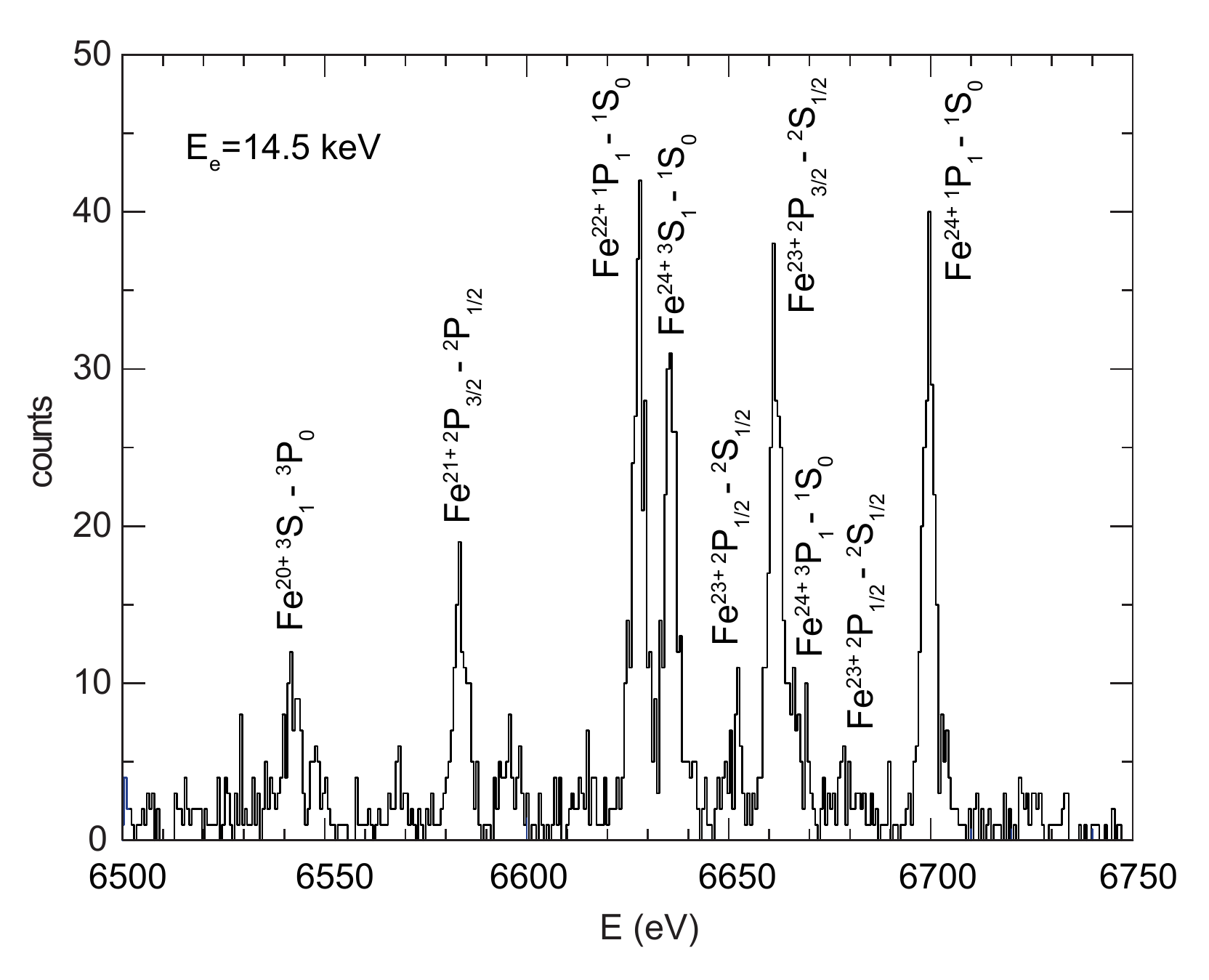}
\caption{Wavelength-dispersive iron X-ray spectrum from DE processes in a Dresden EBIT at 14.5{\u}keV electron beam energy.  \label{fig:fe_kds}}
\end{figure}

\begin{figure}[htbp!]
\centering
\includegraphics[width=0.4\linewidth]{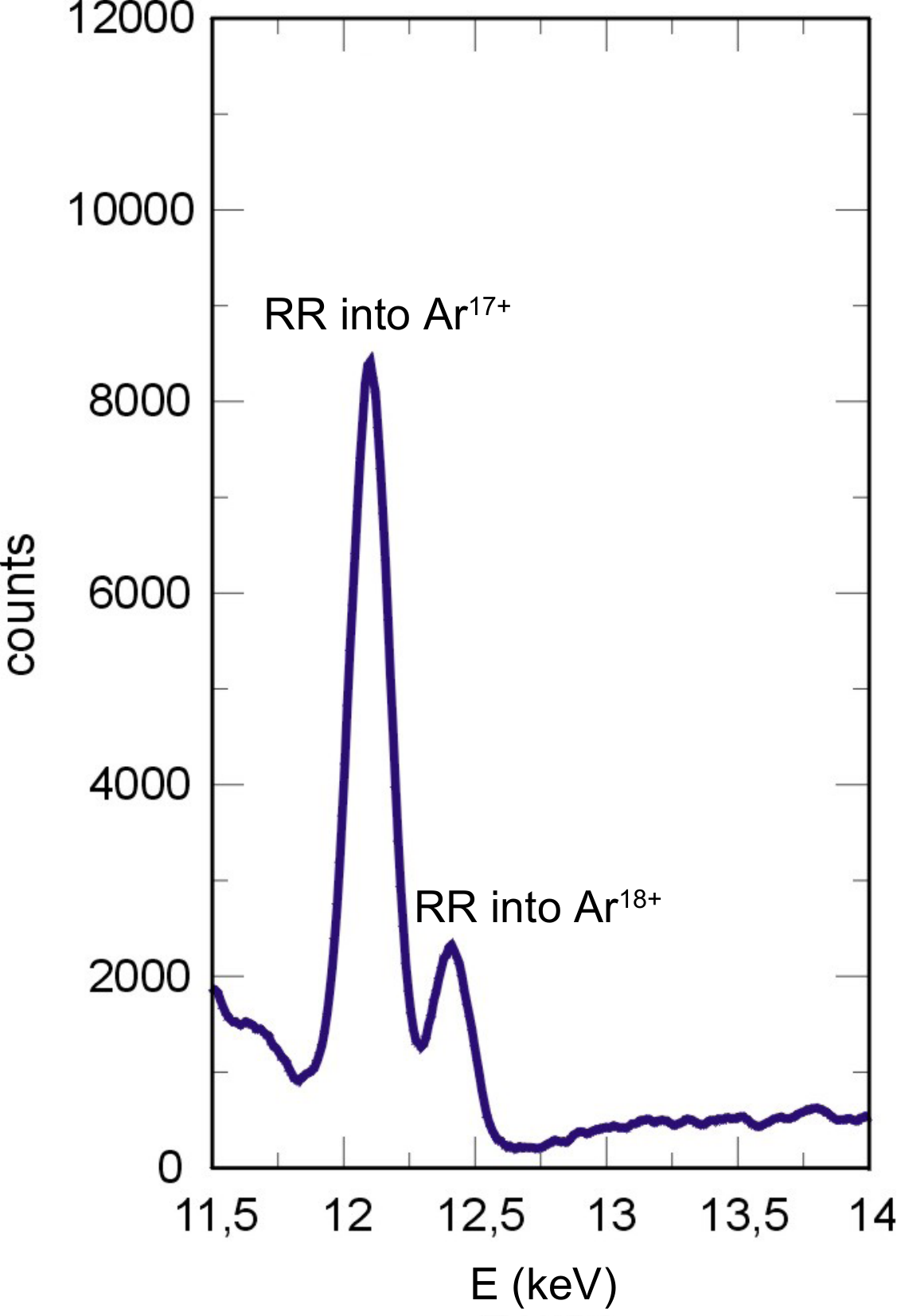}
\caption{Energy-dispersive measured argon RR spectrum from a Dresden EBIT for the electron capture into hydrogen-like and bare argon at an electron beam energy of 8{\u}keV and an electron beam current of 15{\u}mA.  \label{fig:ar-rr-8keV-15mA}}
\end{figure}

In Fig.~\ref{fig:fe_kds} a measured wavelength-dispersive DE spectrum from carbon- to helium-like iron ions is shown. The measurement was realized with a flat crystal and for an electron beam current of 40{\u}mA.
Figure \ref{fig:ar-rr-8keV-15mA} shows an argon RR spectrum as measured with a Dresden EBIT. Here the electron capture into the highest possible argon charge state is shown with a clear signature.

\begin{figure}[htbp!]
\centering
\includegraphics[width=0.65\linewidth]{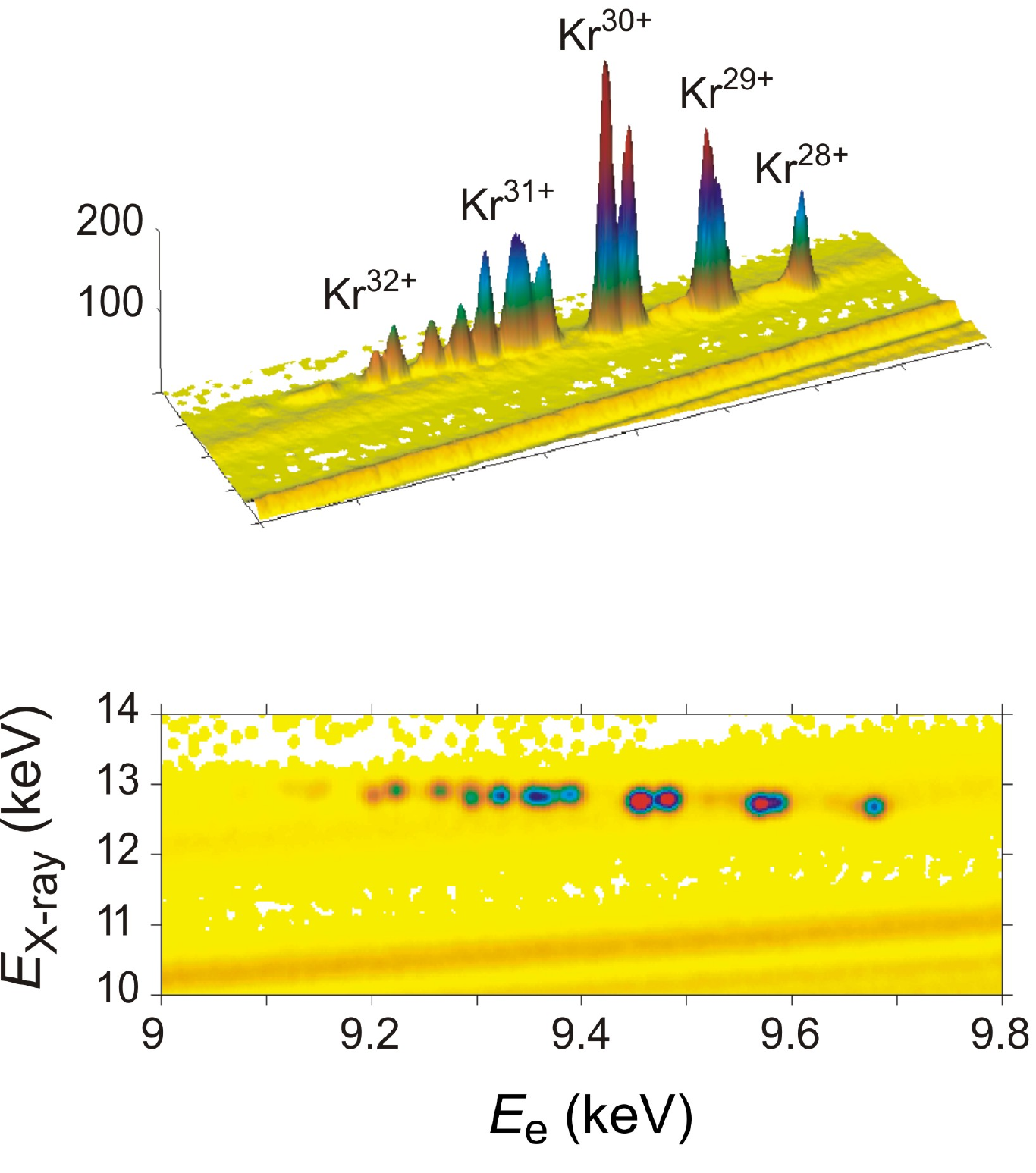}
\caption{Krypton KLL spectrum from DE processes for Kr$^{28+}$ up to Kr$^{32+}$ ions  in a Dresden EBIT  \label{fig:Kr-KLL}}
\end{figure}

DE as well as RR measurements can be realized with  comparatively simple detection techniques. In both cases, without any synchronization with the EBIS, a detector can measure the emitted X-rays passing through a spectroscopic window. More sophisticated are measurements of DE processes where the measurement process must be synchronized with the electron beam energy because DE processes are sharp resonance processes. For example, in Fig.~\ref{fig:Kr-KLL} a krypton DE KLL spectrum for Kr$^{28+}$ up to Kr$^{32+}$ ions is shown.

\begin{figure}[htbp!]
\centering
\includegraphics[width=0.7\linewidth]{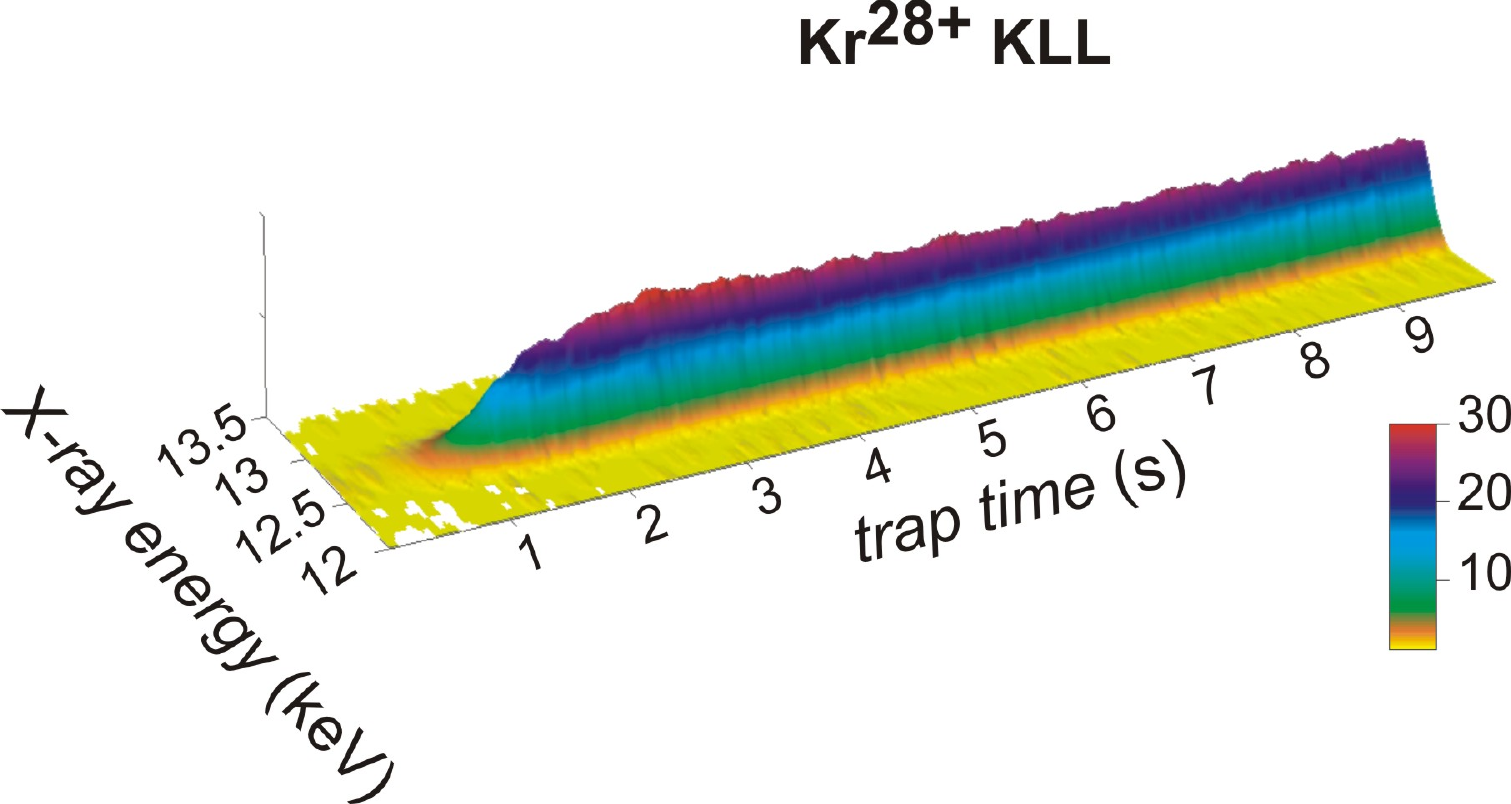}
\caption{Time-resolved krypton DE KLL spectrum for Kr$^{28+}$ measured at a Dresden EBIT  \label{fig:Kr-KLL-t}}
\end{figure}

In addition to the classical energy- or wavelength-dispersive X-ray spectroscopy, time-resolved X-ray spectrometry is possible if we add to each measured count time information about the ionization time in the ion trap (time between closing the ion trap and registration of the X-ray quanta). Such a spectrum is shown in Fig.~\ref{fig:Kr-KLL-t} for a Kr$^{28+}$ DE KLL X-ray transition. The figure shows the evolution of the Kr$^{28+}$ ionization charge state in the ion trap of a Dresden EBIT and demonstrates that such ions can be stored for 10{\u}s or longer.

\section{Applications of EBISs (selected topics)}

As mentioned at the beginning, EBISs are versatile tools for investigations on HCIs in basic and applied research as well as for applications with strong practical relevance. It is not the aim of the present chapter to discuss all these topics in detail, but in the following subsections we will give some examples of the versatile  application potential of HCIs.

\subsection{HCI-FIB for structuring and surface analysis}

Standard focused ion beam (FIB) systems use liquid metal ion sources (LMISs) with their rather small standard spectrum of projectiles. Combining an EBIS, which provides a broader palette of ion species from different elements (from noble gases up to metals) ranging from high to low charge states, with an FIB column leads to new useable physical properties and new applications.

As already mentioned, the potential energy stored in the highly charged ions due to the ionization process leads to power densities of order $10^{12}$ up to $10^{14}${\u}W{\u}cm$^{-2}$ in single ion--surface interactions and therefore to higher yields of secondary ions and secondary electrons per incident ion compared to conventional ions. Preliminary experiments have been done at the CNRS in Marcoussis \cite{bib:Ullmann2007}.

\begin{figure}[htbp!]
\centering
\includegraphics[width=0.7\linewidth]{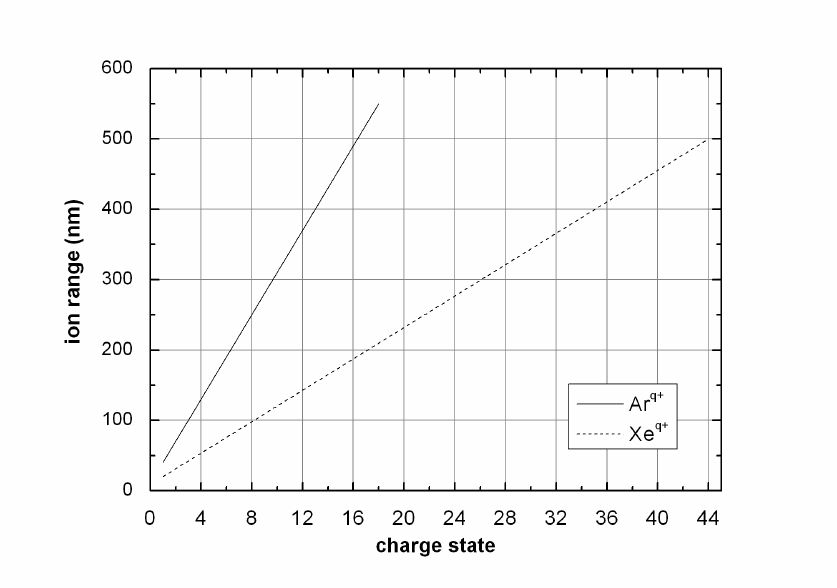}
\caption{Ion range of argon and xenon ions in silicon at different kinetic energies  \label{fig:ion_range}}
\end{figure}

Furthermore, the ion implantation range can be varied by selecting a certain charge state, which is due to the final kinetic energy at a constant acceleration potential (see Fig.~\ref{fig:ion_range}). This can be realized via the integration of a crossed-field ion beam separator, the already introduced Wien filter \cite{bib:Schmidt2009}, which enables the user to separate ions according to their mass-to-charge ratio. Since the ion energy depends on the charge state as well as on the mass, a broad range of projectile energies is available. Also, the intrinsic potential energy of the projectile can be adjusted by the user by selecting the charge state of the ions.

The technology as described above complements the classical spectrum of projectiles for various applications beyond the latest developments using noble gas ions in FIB systems. Considering the ion charge state as a new physical parameter that can be varied offers suitable preconditions in sputtering processes with regard to the produced secondary particles being used for surface analysis (TOF-SIMS). This also takes into account the ability to vary the kinetic energy at a fixed acceleration potential of the ion source and FIB column assembly.

\subsection{Particle therapy}

Cancer is the second most common cause of death worldwide and, statistically, about 33\% of all inhabitants of the European Union will be confronted with some kind of cancer in their lives. Currently about 45\% of all cancer cases can be treated, mainly by surgery and/or by radiation therapy \cite{bib:Auberger2004}.

Within radiotherapy, radiation with hadrons (mainly protons and carbon ions) is the most promising treatment technique.
In contrast to the classical irradiation with X-rays, treatment with ions has significant advantages. Energy deposition within the tissues obeys the Bragg interaction law, i.e.\ the maximum of the energy is deposited at the end of the particle trajectories, hence in the tumour cells. The ion range can be adjusted precisely by varying the kinetic energy of the particles. The intruding ions lose energy inside a dense medium according to the Bragg curve. Most of the kinetic energy is deposited at a certain depth along the ion path. This effect can be used to destroy a tumour while the healthy tissue around is spared (this is important for tumours in the vicinity of risky organs like the medulla).

Another important property of ions as opposed to photons is their electrical charge. Charged particles can be focused and deflected by electrical and magnetic fields, providing the ability to scan with the ion beam over the tumour very precisely. Owing to the variation of the kinetic energy of the ions, it is possible to vary the penetration depth. Hence the tumour can be scanned voxel by voxel (a voxel is a three-dimensional volume element). Besides the electrical properties, the relative biological efficiency (RBE) of the ions  is higher by a factor of about~3. This means that the same dose of carbon ions affects the tumour tissue by a factor of~3 more compared to classical X-rays. A higher RBE allows a lower dose to produce the same effect on the tumour with less damage to the healthy tissue.

Nowadays, a substantial number of medical particle therapy facilities are based on synchrotron accelerators combined with ECR ion sources. Owing to the low purity of the ion beam, electron strippers are required. If using an EBIS ion source and its excellent beam purity, a subsequent electron stripper is no longer necessary. Also, the lower emittance of EBIS sources will lead to a higher transparency of the accelerator structure. For other accelerator strategies, like direct wall accelerators or direct driven accelerators, even the application of room-temperature EBIS ion sources might be possible.

\subsection{Charge breeding}

In nuclear mass measurements involving ions stored in Penning traps, the precision of the experimental method increases proportionally with the charge state of the investigated ions. Furthermore, for proposed nuclear fusion reaction studies, neutron-rich, relatively short-lived radioactive isotopes have to be accelerated to energies above the Coulomb barrier, which requires a small mass-to-charge ratio ($A/q$) to reduce the scale and cost of the accelerator. These and other endeavours rely on the availability of HCI from a broad spectrum of elements or even exotic radioactive species, respectively.

As described above, EBISs or EBITs are HCI sources which have many advantages. However, they are typically loaded via gas injection, which limits the range of available ion species from these machines. The injection of externally produced low charged ions into an EBIS or EBIT and re-extraction as HCIs, known as charge breeding, offers a way to increase the range of available HCI species. Various ion injection techniques have been used for charge breeding. Among these are experimental set-ups including metal vapour vacuum arc and liquid metal or liquid metal ion sources \cite{bib:Nakamura2000,bib:Pikin2006} but also beams of short-lived radioactive isotopes produced using the isotope separation on-line (ISOL) technique \cite{bib:Wenander2010}.

\begin{figure}[htbp!]
\centering
\includegraphics[width=0.7\linewidth]{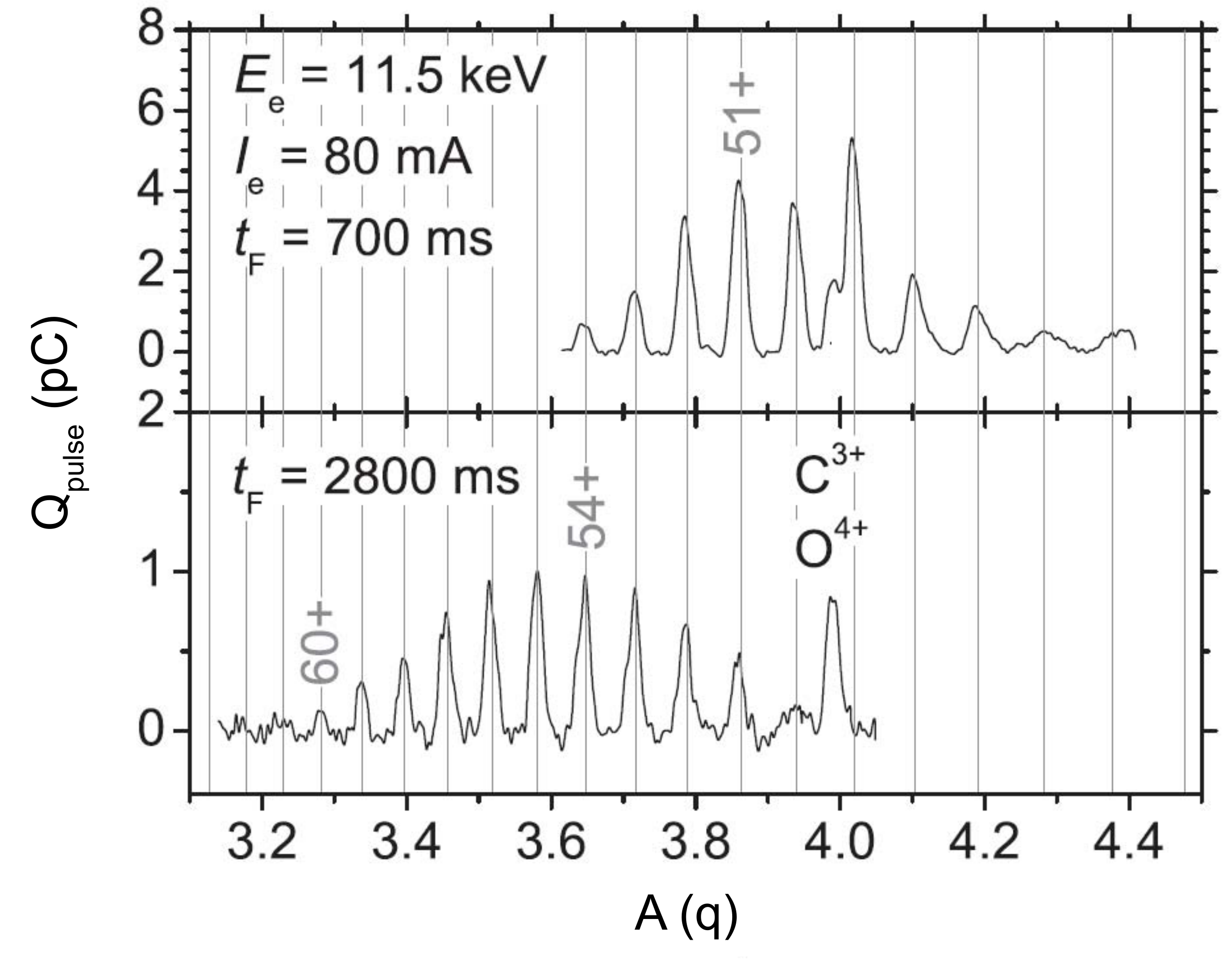}
\caption{Charge breeding spectra with gold ions injected into the trap of an EBIS-A with an external LMIS (from Ref.~\cite{bib:Thorn2012RSI}).  \label{fig:gold60+}}
\end{figure}

Recently, a Dresden EBIS-A was fed by a liquid metal alloy ion source to produce HCI from low charged gold ions. For the charge-bred Au$^{q+}$ ions, charge states up to $q = 60$ were achieved (see Fig.~\ref{fig:gold60+}). This experimental technique represents a clean and elegant way of introducing metals or other elements into the Dresden EBIS-A that cannot be injected directly through a gas valve. Therefore, the range of applications for this type of ion source is extended.

\section{Conclusion}

Highly charged ions (HCIs) are becoming more and more important in basic research as well as technological applications (for example see Refs. \cite{bib:Schmieder1993} --\cite{bib:Shevelko2012}). This is because of their unique properties, which are listed briefly at the beginning of this chapter. In the past, large accelerator devices had to be used for the production of HCIs, while today's sources of HCIs will fit on a table top. This increases the availability of HCIs and opens up new possibilities for a larger community to realize experiments with HCIs.

This chapter has highlighted the electron beam ion source (EBIS) technology as an efficient and elegant way to produce HCIs. The basic physics and working principle of EBISs are explained in detail and compared to other existing HCI sources. Especially, aspects of HCI extraction but also the use of EBISs as sources of X-rays from HCIs trapped inside the machines are discussed.

We hope this work inspires researchers to continue the advancement of this extraordinary ion source technology and creates ideas for new fields of applications of HCIs produced by EBISs in the future.

\end{document}